\documentclass{article}

    \PassOptionsToPackage{numbers, compress}{natbib}

\usepackage{floatrow}
\newfloatcommand{capbtabbox}{table}[][\FBwidth]
\newfloatcommand{capbfigbox}{figure}[][\FBwidth]

\usepackage[final]{neurips_2021}

\usepackage[utf8]{inputenc} 
\usepackage[T1]{fontenc}    
\usepackage{hyperref}     
\usepackage{url}            
\usepackage{booktabs}       
\usepackage{amsfonts}       
\usepackage{nicefrac}       
\usepackage{microtype}      
\usepackage{xcolor}         
\usepackage{natbib}
\usepackage{algorithm}
\usepackage{amsmath}
\usepackage{multirow}
\usepackage[
  separate-uncertainty = true,
  multi-part-units = repeat
]{siunitx}
\usepackage{algpseudocode}

\usepackage{boldline}
\usepackage{enumitem}
\usepackage{setspace}
\usepackage{slantsc}
\usepackage{multirow}
\usepackage{dsfont}
\usepackage{color}
\usepackage{arydshln}
\usepackage{booktabs}
\usepackage{hyperref}
\usepackage{enumitem}
\usepackage{boldline}
\usepackage{enumitem}
\usepackage{setspace}
\usepackage{slantsc}
\usepackage[export]{adjustbox}
\usepackage[lofdepth,lotdepth]{subfig}
\usepackage{xspace}
\usepackage[font=footnotesize]{caption}

\newcommand {\Anurag}[1]{\colornote{red}{Anurag}{#1}}

\title{\frameworkname\,: A Framework for Speech Quality Assessment using Non-Matching References}

\author{%
  Pranay Manocha\thanks{Work done during internship at Facebook Reality Labs Research}\\
  Department of Computer Science\\
  Princeton University\\
  Princeton, NJ \\
  \texttt{pmanocha@cs.princeton.edu} \\
   \And
   Buye Xu \\
   Facebook Reality Labs Research \\
   Redmond, WA \\
   \texttt{xub@fb.com} \\
   \AND
   Anurag Kumar \\
   Facebook Reality Labs Research \\
   Redmond, WA \\
   \texttt{anuragkr@fb.com} \\
}

\newcommand{\ignorethis } [1] {}

\newcommand{\Reals      }     {{\textrm{I\kern-0.18em R}}}

\newcommand{\change     } [1] {\mbox{{\footnotesize $\Delta$} \kern-3pt}#1}

\newlength{\w}

\newcommand{\colornote}[3]{{\color{#1}\bf{#2: #3}\normalfont}}

\newcommand{\Hsimulator} {{{H}}}

\DeclareUnicodeCharacter{2212}{-}
\newcommand{\smallsec}[1]{\noindent {\bf #1.}}

\DeclareMathOperator*{\argmin}{arg\,min}
\newcommand\frameworkname{NORESQA}

\begin{document}

\maketitle

\begin{abstract}
    
   The perceptual task of speech quality assessment (SQA) is a challenging task for machines to do. Objective SQA methods that rely on the availability of the corresponding clean reference have been the primary go-to approaches for SQA. Clearly, these methods fail in real-world scenarios where the ground truth clean references are not available. In recent years, non-intrusive methods that train neural networks to predict ratings or scores have attracted much attention, but they suffer from several shortcomings such as lack of robustness, reliance on labeled data for training and so on. In this work, we propose a new direction for speech quality assessment. Inspired by human's innate ability to compare and assess the quality of speech signals even when they have non-matching contents, we propose a novel framework that predicts a subjective \emph{relative} quality score for the given speech signal with respect to \emph{any provided reference} without using any subjective data. We show that neural networks trained using our framework produce scores that correlate well with subjective mean opinion scores (MOS) and are also competitive to methods such as DNSMOS~\cite{reddy2020dnsmos}, which explicitly relies on MOS from humans for training networks. Moreover, our method also provides a natural way to embed quality-related information in neural networks, which we show is helpful for downstream tasks such as speech enhancement. 
 
\end{abstract}

\section{Introduction}
\vspace{-0.1in}
\label{intro}
 Speech quality assessment is critical for designing and developing a wide range of real-world audio and speech applications, such as, 
 Telephony, VoIP, Hearing Aids, Automatic Speech Recognition, Speech Enhancement etc.
 Clearly, the gold standard for SQA is the evaluation of speech recordings by humans. However, these subjective evaluations are not scalable and can be immensely time-consuming and costly, as they often need to be repeated tens or hundreds of times for every recording. Several objective methods for SQA have been developed to address this problem, PESQ~\cite{rix2001perceptual}, POLQA~\cite{beerends2013perceptual}, VISQOL~\cite{hines2015visqol}, HASQI~\cite{kates2010hearing}, DPAM~\cite{manocha2020differentiable} and CDPAM~\cite{manocha2021cdpam}.
 These methods are \emph{intrusive} or \emph{full-reference} by definition as they are designed to produce a quality score or rating by comparing the corrupted speech signal to its clean reference. 
 However, they suffer from three critical drawbacks.  \emph{First}, the requirement of a paired clean reference for quality assessment limits their applicability to real-world scenarios as the paired clean reference is likely not available in those cases. \emph{Second}, these methods have acknowledged shortcomings such as sensitivity to perceptually invariant transformations~\cite{hines2013robustness}, therefore hindering stability in more diverse tasks such as speech enhancement. \emph{Lastly}, these metrics are non-differentiable and cannot be directly leveraged as training objectives in the context of neural networks.

To address these problems, a recent trend has been to develop \emph{non-intrusive}~\cite{loizou2011speech} methods using neural networks~\cite{reddy2020dnsmos,soni2016novel,spille2018predicting,fu2018quality,andersen2018nonintrusive,lo2019mosnet,gamper2019intrusive,avila2019non,yu2021metricnet,dong2020attention,zhang2021end,patton2016automos}. In most cases, the primary approach is to train a model to predict objective (e.g., PESQ) and/or subjective (e.g., MOS) scores. However, generalization to unseen perturbations and tasks remains a concern~\cite{dong2019classification}, and most methods have not found wide-spread uses for SQA. Given that matching human subjective ratings is the ultimate motivation, some recent works try to train neural networks directly on MOS scores~\cite{patton2016automos,lo2019mosnet,reddy2020dnsmos}. DNSMOS~\cite{reddy2020dnsmos}, in particular, trains neural networks on a very large-scale MOS database. However, collecting such a dataset is an uphill task and requires considerable resources. 
For e.g., one requires uniformity with respect to hardware (headphones/speakers), listening environments etc., among hundreds and thousands of raters to ensure that ratings are consistent. Otherwise, a significant amount of noise can creep into the dataset, making it unreliable for training. In DNSMOS \cite{reddy2020dnsmos}, almost half of the recordings have a standard deviation of more than 1 (MOS $\pm$1 ), which poses challenges in  training robust models due to noisy labels.

Lastly, we would like to point out an inherent challenge of the conventional formulation of any non-intrusive metric. The problem might lie in the \emph{lack of reference} itself. While training any model on subjective scores, we expect the model to implicitly learn the distribution of references that are consciously or unconsciously used by human listeners. These references can be highly varied and strongly influenced by each individual's past experience and even the mood when participating in the evaluations. Learning such a distribution can be an extremely hard problem, especially when no specific constraints on the distribution of clean reference are provided to the model during training.

In this work, we propose a novel and alternate framework for speech quality assessment. Instead of a completely reference-free approach, our framework relies on random non-matching references (NMRs) of known qualities, and is designed to provide a relative assessment of speech quality with respect to the NMRs. The inspiration for the approach comes from human's ability to do the same. Given two completely random speech recordings, it is highly likely that a human would be able to compare them with respect to quality irrespective of the actual speech content. This innate ability to compare the quality of two speech recordings in an ``unsupervised" setting (or non-matching conditions) holds true even when the recordings contain different speakers, languages, words, and so on. Moreover, comparative tests are relatively easier for humans than absolute rating, and relative scores tend to have lower variance and less noise. 

\looseness=-1
Motivated by the above points, we propose~\frameworkname\, - \emph{NOn-matching REference based Speech Quality Assessment}. Within this framework, we propose to learn neural networks that can predict a \emph{relative quality score for a given speech recording with respect to any provided reference}. A few key characteristics of~\frameworkname\, are:~(i) unlike full-reference objective metrics, it remains usable in real world situations by relying on NMRs which are readily available;~(ii) it addresses the problem of lack of reference in non-intrusive methods by providing an NMR, thereby providing the necessary grounding for the model to learn and predict acoustic quality differences; ~(iii) pairwise comparisons are expected to have lower variance and less noise than absolute ratings (e.g MOS); and~(iv) any learning within this framework is ``unsupervised'' in the sense that we do not require any manually labeled dataset for training models.

To summarize the \emph{key contributions} of the paper:~(1) we propose a novel framework for speech quality assessment that relies on NMRs;~(2) we propose methods to train neural networks within this framework using multi-task multi-objective learning; and (3) we evaluate our framework comprehensively through several subjective and objective evaluations as well as exploring its utility in a downstream task like speech enhancement.


\section{Related Work}
\label{sec:format}
\vspace{-0.1in}
\smallsec{Non-Intrusive Methods}
\label{related_works}
Some of the earliest non-intrusive methods were based on complex hand-crafted, rule-based systems~\cite{malfait2006p,grancharov2006low,narwaria2011nonintrusive,sharma2014non}. 
Although they are automatic and interpretable, they tend to be task-specific, and do not generalize well. Moreover, these methods are non-differentiable which limits their uses within deep learning frameworks. To overcome the last concern, various neural network based methods have been developed~\cite{reddy2020dnsmos,soni2016novel,spille2018predicting,fu2018quality,andersen2018nonintrusive,lo2019mosnet,gamper2019intrusive,avila2019non,yu2021metricnet,dong2020attention}. However, the issue of task-specificity and generalization remained. To overcome this, researchers proposed to train models directly on a dataset of human judgment scores~\cite{reddy2020dnsmos,lo2019mosnet,patton2016automos}. Reddy et al.~\cite{reddy2020dnsmos} used a multi-stage self-teaching model~\cite{kumar2020sequential} to learn quality in the presence of noisy ratings. Nonetheless, non-intrusive metrics lag behind intrusive metrics in terms of correlation to human listening evaluations and adoption in practical cases. 

\smallsec{Vision}
Researchers in the vision community have also explored the field of no-reference image quality assessment. Various works have looked at approaches using learning to rank~\cite{liu2017rankiqa,zhang2019ranksrgan,lin2018hallucinated} rather than training on absolute ratings. However, to the best of our knowledge, no prior work has explored a framework that relies on random NMRs to provide a relative assessment of quality.

\smallsec{Multi-task learning for audio quality}
Multi-task learning (MTL)~\cite{caruana1997multitask} has been beneficial to many speech applications~\cite{seltzer2013multi,chen2015multitask,fu2016snr,fakoor2018constrained}.
MTL aims to leverage useful information contained in multiple related tasks to help improve the generalization performance on all tasks.
Serra et al.~\cite{serra2020sesqa} proposed SESQA learned from various objectives like MOS, pairwise ranking, score consistency, and other metrics like PESQ and STOI etc,. However, their approach uses paired clean and noisy recordings for training and also requires access to a dataset of subjective ratings.
In contrast, our proposed quality assessment framework works in a ``unsupervised'' setting and does not require any labeled data. We leverage MTL to learn a preference and a quantification task for quality assessment. 

\begin{figure}[t!]
\vspace{-0.3in}
\centering
\setlength{\w}{0.32\columnwidth}
\setlength{\tabcolsep}{2pt}
\begin{tabular}{cc}
\includegraphics[width=0.30\columnwidth]{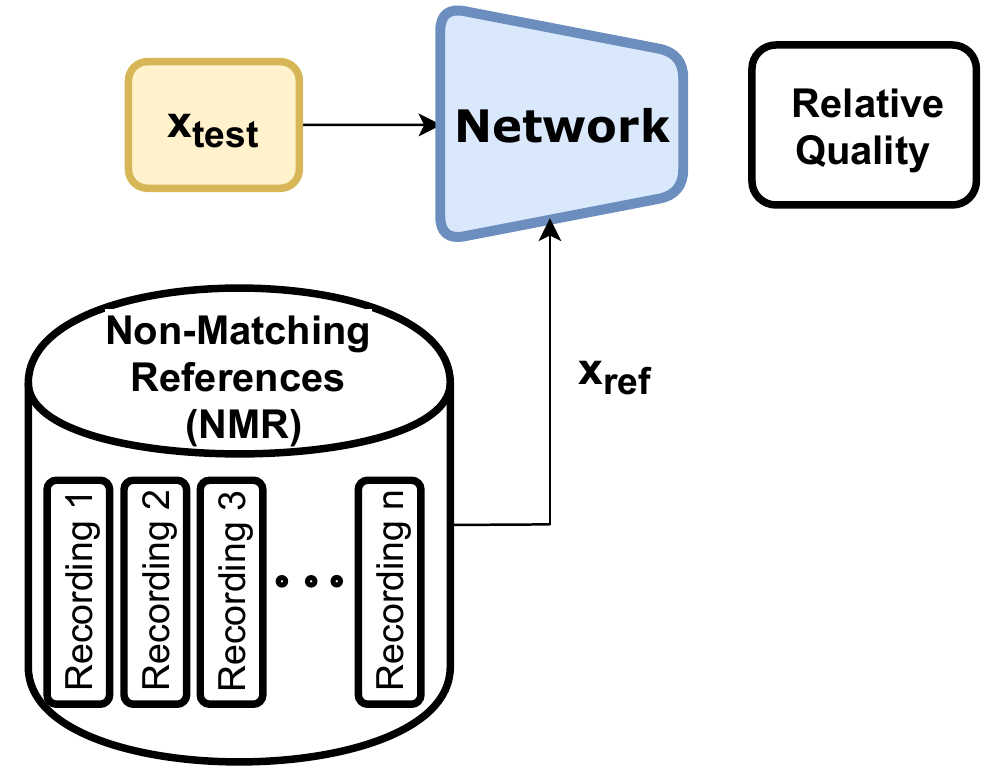} &
\hspace{0.001in}
\includegraphics[width=0.65\columnwidth]{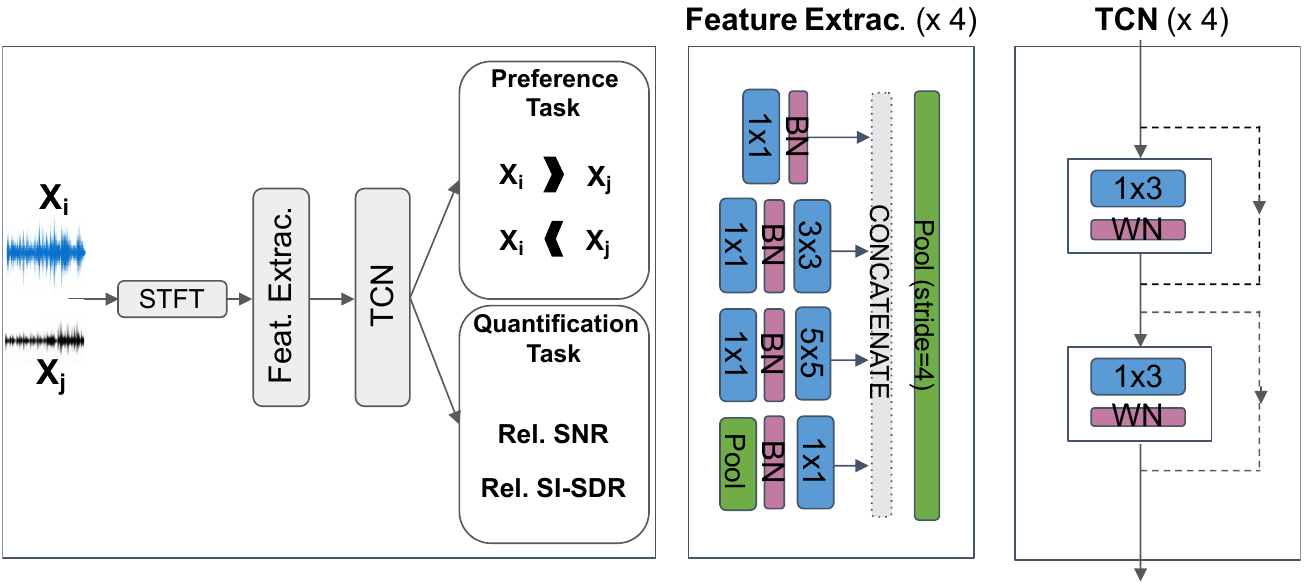} \\
\begin{minipage}{0.001\columnwidth}

{\footnotesize {(a)}}
\end{minipage} &
\begin{minipage}{0.75\columnwidth}
\centering
{\footnotesize {(b)}}
\end{minipage}
\end{tabular}
\vspace{-0.15in}
\caption{ \textit{Left:}(a) {\bf Proposed \frameworkname\, framework} takes a test recording, and a randomly chosen NMR from a set of NMRs and predicts the relative quality. \textit{Right}: {\bf \frameworkname\, Architecture}~(b) Our network takes two non-matched recordings ($x_\mathtt{i}$ and $x_\mathtt{j}$) and outputs:~(i) which recording is cleaner (\textit{preference-task}); and~(ii) relative quality difference using SI-SDR and SNR (\textit{quantitative-task}).\vspace{-1\baselineskip}}
\label{model_diagram}
\end{figure}

\vspace{-0.1in}
\section{\frameworkname\, Framework}
\label{sec:framework}
\vspace{-0.1in}
\looseness=-1
Our framework,~\frameworkname\,, is designed to assess the quality of a given speech recording using Non-Matching References (NMRs). This novel framework is a \emph{generic idea}, and within it, one can develop a variety of methods to produce a relative quality score (referred to as~\frameworkname\, score) for a test input, $x_{test}$, with respect to any given reference, $x_{ref}$. We propose a deep neural network for~\frameworkname\, and represent it by the function $\text{\frameworkname} = \mathcal{N}(x_{test},x_{ref})$. Given that we do not rely on any human-labeled data, the crucial components of the framework include designing tasks and objective functions that can help learn a quality score. This is along the lines of self-supervised learning ~\cite{le2020contrastive, devlin2018bert, baevski2020wav2vec}, where supervised tasks are designed on unlabeled data for training neural networks. 

\vspace{-0.1in}
\subsection{Overview}
\label{sec:overview}
\vspace{-0.05in}
\looseness=-1
The~\frameworkname\, framework takes in two recordings as inputs - test recording $x_\mathtt{test}$ and another randomly chosen recording $x_\mathtt{ref}$. Fig~\ref{model_diagram}(a) is a simple illustration of the framework. 
In our current approach,~\frameworkname\, has the following properties:~(i) Non-negative (by design) $\mathcal{N}(x_{test}, x_{ref}) \geq 0$;~(ii) Monotonic (by design) if $\mathcal{N}(x_{test 1}, x_{ref}) \geq \mathcal{N}(x_{test 2}, x_{ref})$, then $\mathtt{m}(x_{test1}) \leq \mathtt{m}(x_{test2})$, where $\mathtt{m}$ is any quality assessment measure as defined in Section~\ref{sec:mtl};~(iii) a way to predict ``sign", which helps differentiate $\mathcal{N}(x_a, x_b)$ vs. $\mathcal{N}(x_b, x_a)$.  We do not enforce other metric properties~\cite{li2004similarity,chen2009similarity} to allow flexibility in defining tasks and objectives for training the neural networks. Moreover, even human judgment of similarity may not constitute a metric~\cite{tversky1977features}, and hence there is no pertinent reason which necessitates~\frameworkname\, to have metric properties. 

\looseness=-1
The network architecture is shown in Fig~\ref{model_diagram}(b). The neural network model is designed to detect and quantify degradation at the frame-level (frame $\approx$ 32 ms of audio). To learn quality, we propose a multi-task multi-objective training mechanism in which the network aims to classify which input is better, as well as ``quantify by how much", in terms of two measures, scale-invariant signal to distortion ratio (SI-SDR) and signal-to-noise ratio (SNR).
Additionally, our framework outputs frame-level judgments, by which we can detect and quantify which frames are degraded in quality with respect to the NMRs.


\vspace{-0.1in}
\subsection{Architecture}
\label{sec:architecure}
\vspace{-0.05in}
\looseness=-1
\frameworkname's architecture (shown in Fig~\ref{model_diagram}(b)) comprises of three key components: a \emph{feature-extraction} block, a \emph{temporal-learning} block and \emph{task-specific} heads. The feature-extraction block learns representations while maintaining the temporal structure of the signal. 
These representations are then fed to the temporal-learning block to learn long-term dependencies using temporal convolutional networks (TCN). The resulting embeddings are then fed to the task-specific heads to produce frame-wise outputs. Note that the architecture is an instance of shared parameters for the two inputs, in other words both inputs are processed by the same network. 

\looseness=-1
\smallsec{Feature-Extraction Block}
 The inputs to the network are 3-second audio excerpts, represented by their Short-Time Fourier Transform (STFT). We stack together the magnitude and phase of the STFT as two channels, leading to a $2 \times T \times F$ dimensional representation of every audio recording where $T$ is the number of frames and $F$ is number of frequency bins. More specifically, STFTs are computed using a hamming window of size $32$ ms with 50\% overlap. Only the 256 positive frequencies without the zeroth bin are used. We use the Inception~\cite{szegedy2015going} architecture in the feature extraction block. Please refer to the supplementary material for details of the feature-extraction block.
 
\looseness=-1
\smallsec{Temporal Learning Block}
The temporal learning block is aimed at learning long-term temporal information from frame-level representations generated by the feature-extraction block. We use Temporal Convolutional Networks (TCNs)~\cite{bai2018empirical} in this block. While the phonetic speech content can change per frame, the acoustics (recording conditions, background noise, distortions, etc) necessitates capturing long-term history for quality assessment. A cascade of TCNs offers a flexible way to learn both short-term (local) and long-term information. We request the readers to refer to supplementary material for precise details of the TCNs used in this block. 

The learned weights across the first two blocks are shared between the two inputs to our model. The outputs of the temporal learning block for both inputs are concatenated, and then fed into each of the task-specific heads.
Next, we describe the multi-task multi-objective learning mechanism:

\subsection{Multi-task and multi-objective learning}
\label{sec:mtl}
\looseness=-1
The~\frameworkname\, network architecture described above is trained through a multi-task multi-objective approach. Since we do not have any perceptual labels, we rely on two signal processing measures, Signal-to-Noise Ratio (SNR) and Scale-Invariant Signal to Distortion Ratio (SI-SDR) to compare the quality of the two inputs.

\looseness=-1
SNR is measured as the ratio of the signal power to the noise power and is primarily meant only for additive noises. Consider a mixture signal $x$, $x = s + n \in \mathbb{R}^L$ where $s$ is the clean signal and $n$ is the noise signal, then
\begin{equation}
\mbox{SNR} = 10\log_{10} \left(\frac{||s||^2}{||s - x||^2}\right)
\label{snr}
\end{equation}
$10log_{10} ()$ factor measures SNR in dB-scale, and a higher SNR implies better signal quality. Yuan et al.~\cite{yuan2019signal} also showed that SNR as a distance metric had better properties than conventional metrics (like Euclidean distance).

SI-SDR is a measure that was introduced to evaluate performance of speech processing algorithms. It is invariant to the scale of the processed signal and can be used to quantify quality in diverse cases, including additive background noises as well as other distortions~\cite{le2019sdr}. 

Given a noisy mixture $x$ and it's clean counterpart $s$, the SI-SDR is defined as:
\begin{equation}
\mbox{SI-SDR} = 10\log_{10} \left(\frac{||\alpha s||^2}{||\alpha s - x||^2}\right), \,\mbox{where}\, \alpha = \argmin_{\alpha} ||\alpha s - x|| =  \frac{x^Ts}{||s||^2}
\label{sdr}
\end{equation}
This property of scale invariance is very important since both speech quality and intelligibility to a large extent are invariant to scaling~\cite{moore2012introduction}. SI-SDR has been shown to be an effective approach for quality measurement~\cite{kolbaek2020loss,sivaraman2020self}.
\looseness=-1
The motivation behind relying on SNR and SI-SDR is that these are two of the most fundamental and generic measures to quantify the quality of a signal. They have been used extensively in training and evaluating audio source separation and speech enhancement algorithms~\cite{tan2020sagrnn,luo2019conv,hu2007evaluation,kolbaek2020loss,sivaraman2020self}, and also correlate well with human perception on various realistic tasks~\cite{mizumachi2013relationship,ma2011snr}. 

We use the SNR and SI-SDR measures as ``quality labels" for all speech recordings, and the objective of the network is to solve two tasks. Given two audio inputs, the network solves:~(a) a \emph{Preference Task}: which input audio is of better quality?;~and~(b) a \emph{Quantification Task}: what is the quality difference between the two audio inputs w.r.t their ``quality labels"?
Separating the learning task into two separate tasks is important is because the inputs to our model are non-matching pairs. If we use signed intervals, it might be non-trivial for the model to learn to discard speech content differences and learn quality features. Disentangling the learning through two separate tasks makes it easier for the mode to focus on quality attributes. It also makes the model easier to use - in the sense that if only preference results are needed or if only absolute quality difference is required, the appropriate head can be used while discarding the unused one. Lastly, the framework can be easily extended through the separate preference and quantification tasks. For example, we can have more than 2 inputs and the preference task can be designed to predict a ranked order in terms of quality. 

Note that both of these tasks are relative assessments as opposed to absolute quality measurements. Human listeners find relative assessment an easier task than absolute quality ratings~\cite{teodorescu2016absolutely} and relative subjective scores are typically more robust than absolute scores~\cite{raake2012talk}. Relative scores tend to give more repeatable results~\cite{rix2006objective} and are likely to have lower variance and bias. Hence, we argue that neural networks trained on these relative assessments will likely lead to better generalization. Moreover, just as humans can compare the quality between non-matching recordings, our training mechanism also follows the same strategy.

\looseness=-1
\smallsec{Preference Task} The preference task is formulated as a binary classification problem. Let $\mathbf{x}_{ij} = (x_i, x_j)$ be an \emph{ordered} pair input to the network, with $x_i$ as first input and $x_j$ as second input. Let $sdr_{x_i}$ and $sdr_{x_j}$ be the SI-SDRs of $x_i$ and $x_j$ respectively. The label $\mathbf{y}_{ij}$ for $\mathbf{x}_{ij}$ is a 2 dimensional, one-hot vector, with $\mathbf{y}_{ij} = [1, 0]$ if $sdr_{x_i} > sdr_{x_j}$, otherwise $\mathbf{y}_{ij} = [0, 1]$. The preference-task head is a small, fully convolutional network and outputs a distribution of which input is cleaner at frame-level. The frame-level distributions are temporally averaged to produce the recording level distribution $\mathbf{p}_{ij}$ which are then used in the standard cross-entropy function to compute loss for this task. 
\begin{equation}
\label{eq:pref_task}
L_P(\mathbf{x}_{ij}, \mathbf{y}_{ij}) = \sum_{k=1}^{2}{-y^k_{ij}\log(p^k_{ij})}
\end{equation}

\looseness=-1
\smallsec{Quantification Task} This task is designed to quantify the quality differences between $x_i$ and $x_j$. Let $snr_{x_i}$ and $snr_{x_j}$ be SNRs of $x_i$ and $x_j$ respectively. This task consists of two output heads, one for predicting quality differences in terms of SNR, $\MakeUppercase{\Delta} snr_{ij} = |snr_{x_i} - snr_{x_j}|$, and the other for SI-SDR $\MakeUppercase{\Delta}sdr_{ij} = |sdr_{x_i} - sdr_{x_j}|$. Architecturally both heads are identical, the details of which are provided in the supplementary material. Trivially, one can formulate $\Delta snr$ or $\Delta sdr$ prediction as regression problems~\cite{lo2019mosnet,serra2020sesqa}. Here, we formulate them as classification problems. By applying tricks from deep neural network based classification methods, we are able to learn much more robust models. 

Let $\MakeUppercase{\Delta}sdr_{max} = \max_{ij} |sdr_{x_i} - sdr_{x_j}|$ be the maximum absolute difference in SI-SDR among all pairs of $x_i$ and $x_j$ in the dataset. We divide the range of $\MakeUppercase{\Delta}sdr$, from $0$ to $\MakeUppercase{\Delta}sdr_{max}$, into $K$ equally spaced bins. Each bin is a class and the label for $\mathbf{z}_{ij}$ for input $\mathbf{x}_{ij}$ is $k^{th}$ class if $\MakeUppercase{\Delta}sdr_{ij}$ lies in the range $\left[\frac{(k-1)\MakeUppercase{\Delta}sdr_{max}}{K}, \frac{k \cdot \MakeUppercase{\Delta}sdr_{max}}{K}\right]$. The output of the SI-SDR head of the network is a probability distribution over all $K$ classes. Similar to the Preference task, the network first produces frame-level distributions which are then temporally averaged to obtain the output $\mathbf{d}_{ij}$ for $\mathbf{x}_{ij}$. 

The $K$ classes here are not entirely independent and share strong inter-class relationships. For example, the $k^{th}$ class is semantically much closer to $(k+1)^{th}$ class than, say $(k+4)^{th}$ class.
 This raises concerns over using the standard cross-entropy loss function with one-hot vectors as target labels. Moreover, training the network with the standard cross-entropy loss can lead to overconfident prediction and a lack of generalizations in unseen conditions~\cite{muller2019does}. This is especially a concern for us as we expect the network to learn the quality differences between non-matching inputs. Considering these factors, we propose to use \emph{gaussian-smoothed-labels}~\cite{geng2016label} for computing the loss function. Label smoothing discounts certain probability mass from the true label and redistributes it to others. Formally, let $v$ be the class-index of the true class. The smoothed labels, $^{s}\mathbf{z}_{ij}$ are obtained as

\begin{equation}
^{s}\mathbf{z}_{ij}^k = \begin{cases}
0.6 &\text{k=v}\\
0.2 &\text{k=v$\pm$1}\\
0 &\text{otherwise}\\
\end{cases}
\end{equation}

\looseness=-1
The same procedure is applied to the SNR head as well. The overall objective loss function for the quantification task is then defined as
\setlength{\abovedisplayskip}{1pt}
\setlength{\belowdisplayskip}{1pt}
\begin{equation}
\label{eq:quant_loss}
L_{Q}(\mathbf{x}_{ij}, ^{s}\hspace{-0.025in}\mathbf{z}_{ij}) = L_{sdr}(\mathbf{x}_{ij}, ^{s}\hspace{-0.025in}\mathbf{z}_{ij}) + L_{snr}(\mathbf{x}_{ij}, ^{s}\hspace{-0.025in}\mathbf{u}_{ij}) =-\sum_{k=1}^{K} \, ^{s}\hspace{-0.025in}z_{ij}^{k}\log(d_{ij}^{k}) - \sum_{k=1}^{K} \, ^{s}\hspace{-0.025in}u_{ij}^{k}\log(t_{ij}^{k})
\end{equation}
Similar to $^{s}\mathbf{z}_{ij}$ and $\mathbf{d}_{ij}$, $^{s}\mathbf{u}_{ij}$ and $\mathbf{t}_{ij}$ are the smoothed-label targets and probability outputs, respectively, for the SNR head. The total loss for training is $L = L_{P} + L_{Q}$.

\subsection{Training Procedure}
\label{sec:train_mech}
\looseness=-1
We now describe our training procedure. We assume the availability of a clean speech database $\mathcal{D}_c$. The training input $\mathbf{x}_{ij}$ is created by sampling two clean recordings $s_i$ and $s_j$ from $\mathcal{D}_c$. $s_{i}$ and $s_{j}$ are corrupted to produce $x_{i}$ and $x_{j}$. The degradation's we use can be largely grouped under two categories~(a) additive noise degradations, and~(b) speech distortions based on signal manipulations - \emph{Clipping}, \emph{Frequency Masking}, and \emph{Mu-law compression}. For additive noise, we sample noise recordings, $n_i$ and $n_j$, from a noise database and add them to $s_i$ and $s_j$ at SNR levels uniformly sampled from the range -15dB to 60dB. This corresponds to a range of -15dB to +25dB in terms of SI-SDR. To ensure that SNR is consistent for $x_i$ and $x_j$, and for more stable training, the noise recordings $n_i$ and $n_j$ should generally belong to similar noise types.
For speech distortions, we use the Audiomentations~\footnote{\footnotesize{ \url{https://github.com/iver56/audiomentations}}} toolkit. It allows one to select levels for various distortions and we select levels that always lie in the range of -15dB to +25dB. Once we have the degraded signals ($x_i$ and $x_j$) and their SNR and SI-SDR ($snr_{x_i}$, $snr_{x_j}$, $sdr_{x_i}$, $sdr_{x_j}$), we can train the network as described in previous sections. Note that, for degradations under the second category, the \emph{quantification-task} loss $L_Q$, includes loss only from the SI-SDR head, $L_{sdr}$. SNR cannot be accurately computed for these distortions (residual $x-s$ may not be orthogonal to $s$) and hence we rely only on SI-SDR for training the network in these cases.

\subsection{Usage}
\label{sec:usage}
\looseness=-1
Once the network is trained, we can predict the quality of a test input $x_{test}$ with respect to any provided reference $x_{ref}$. As mentioned earlier, this reference need \textbf{not} be the matching clean reference. The quality score,~\frameworkname\,, of $x_{test}$ w.r.t $x_{ref}$ is obtained from the SI-SDR head of the network. It is obtained as 
\begin{equation}
\label{sdr}
\text{\frameworkname}_{x_{test}, x_{ref}} = \sum_{k=1}^{K} d_{x_{test}, x_{ref}}^k \mu^k,~~\text{where} ~~ \mu^k = \frac{1}{2} \left[\frac{(k-1)\MakeUppercase{\Delta}sdr_{max}}{K} +  \frac{k \cdot \MakeUppercase{\Delta}sdr_{max}}{K}\right]
\end{equation}
Here $d_{x_{test}, x_{ref}}^k$ and $\mu^k$ are the probability and the mean of the $\Delta sdr$ range of the $k^{th}$ class respectively. The~\frameworkname\, score gives us only the magnitude of difference between $x_{test}$ and $x_{ref}$. The output of the preference-task tells us the ``sign", whether $x_{test}$ is better or worse than $x_{ref}$. 

Our method provides a way to compare the quality of any two given speech recordings. However, in practice, one might often be interested in measuring ``true" or ``absolute'' quality, which under our framework can be done by using clean or high-quality speech recordings as references. More specifically, the $x_{ref}$ samples should come from \emph{any} clean speech database. Moreover, to reduce variance in the estimate, we can sample multiple references and obtain an average~\frameworkname\, score, $\text{\frameworkname}^{avg}_{x_{test}, x_{ref}} = \frac{1}{n} \sum_{i=1}^n \text{\frameworkname}_{x_{test}, x_{ref}^i}$, where
$x_{ref}$ = \{$x_{ref}^i$\} $\forall i \in [1,n]$ are $n$ NMRs sampled from the database.

\vspace{-0.1in}
\section{Experimental Setup}
\label{sec:setup}
\vspace{-0.1in}

\smallsec{Datasets, Implementations and Baselines} For training and validation, we choose the clean speech recordings from the DNS Challenge~\cite{reddy2020interspeech}. FSDK50~\cite{fonseca2020fsd50k} serves as the noise dataset for additive noise degradations. Along with additive noise, clipping and frequency masking distortions are used during training. For robustness and better generalization to realistic conditions, we also add reverberation using room impulse responses from DNS Challenge dataset. For the test-set, we use TIMIT~\cite{garofolo1993darpa} as the source for clean speech, and ESC-50~\cite{piczak2015esc} dataset for noise recordings. The test-set also includes Gaussian noise addition and Mu-law compression as unseen degradations. 

Our method was implemented using Pytorch~\cite{paszke2019pytorch}. We use Adam~\cite{kingma2014adam} optimizer with a learning rate of $10^{−4}$ with a batch size of 32. We train the network for 1000 epochs using 4 Tesla V100 gpus. Please refer to supplementary material for further details on the datasets and implementations.

\looseness=-1
We use the well-known PESQ~\cite{rix2001perceptual} and CDPAM~\cite{manocha2021cdpam} as full-reference baselines for SQA. CDPAM is a full-reference neural network based approach trained on human-pairwise judgements. Being full-reference, these methods require the paired (matching) clean reference for quality assessment. We use DNSMOS~\cite{reddy2020dnsmos} as non-intrusive SQA method. It is trained on a large scale dataset of MOS ratings. However, no such large-scale dataset of MOS ratings is publicly available.

\looseness=-1
\smallsec{Evaluation Methods} We exhaustively evaluate our proposed framework in a variety of ways. The first series of evaluations, which we call \emph{Objective Evaluations} for simplicity, focus on evaluating the models performance on each task (preference and quantification), and its in-variance to different NMR conditions, and so on. The second series of evaluations, referred to as \emph{Subjective Evaluations}, applies our model to a series of speech problems and evaluates how well~\frameworkname\, is correlated to subjective ratings (e.g MOS). Lastly, we show the utility of our model in improving learnability in a downstream task, speech enhancement.

\section{Results}
\label{sec:results}
\subsection{Objective Evaluations}
\label{sec:obj_val}
\smallsec{Performance on Preference and Quantification Tasks}
The test inputs are created by randomly selecting two different clean recordings from the test set (TIMIT) and then degrading them as per the procedure in Sec.~\ref{sec:train_mech}. We then measure the performance of the model on each task. On the Preference task, the model achieves an accuracy of \textit{97.3\%}.

\begin{figure}[t!]
\vspace{-0.24in}
\centering
\setlength{\w}{0.33\columnwidth}
\setlength{\tabcolsep}{0pt}
\begin{tabular}{ccc}
\includegraphics[width=\w]{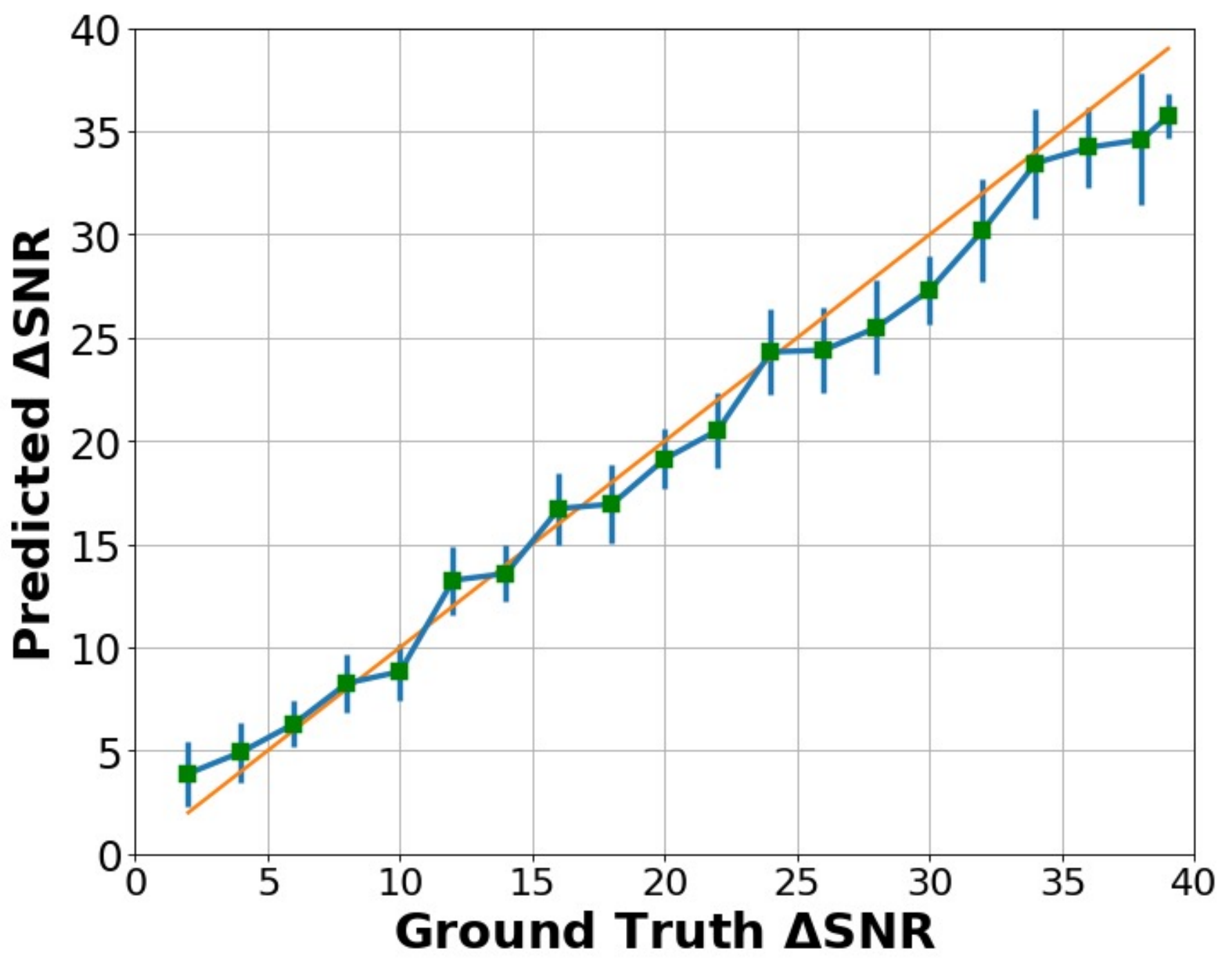} &
\includegraphics[width=\w]{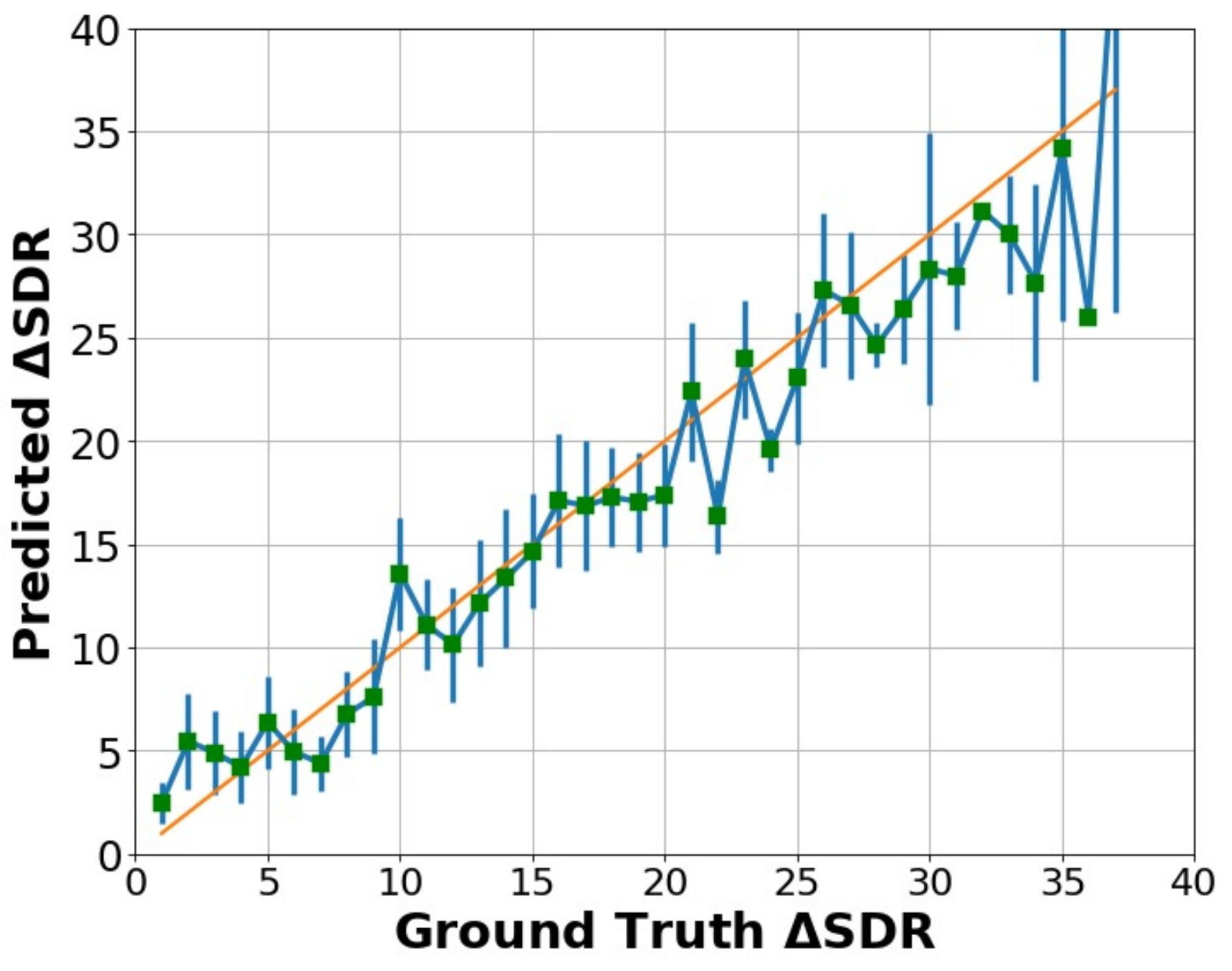} &
\includegraphics[width=\w]{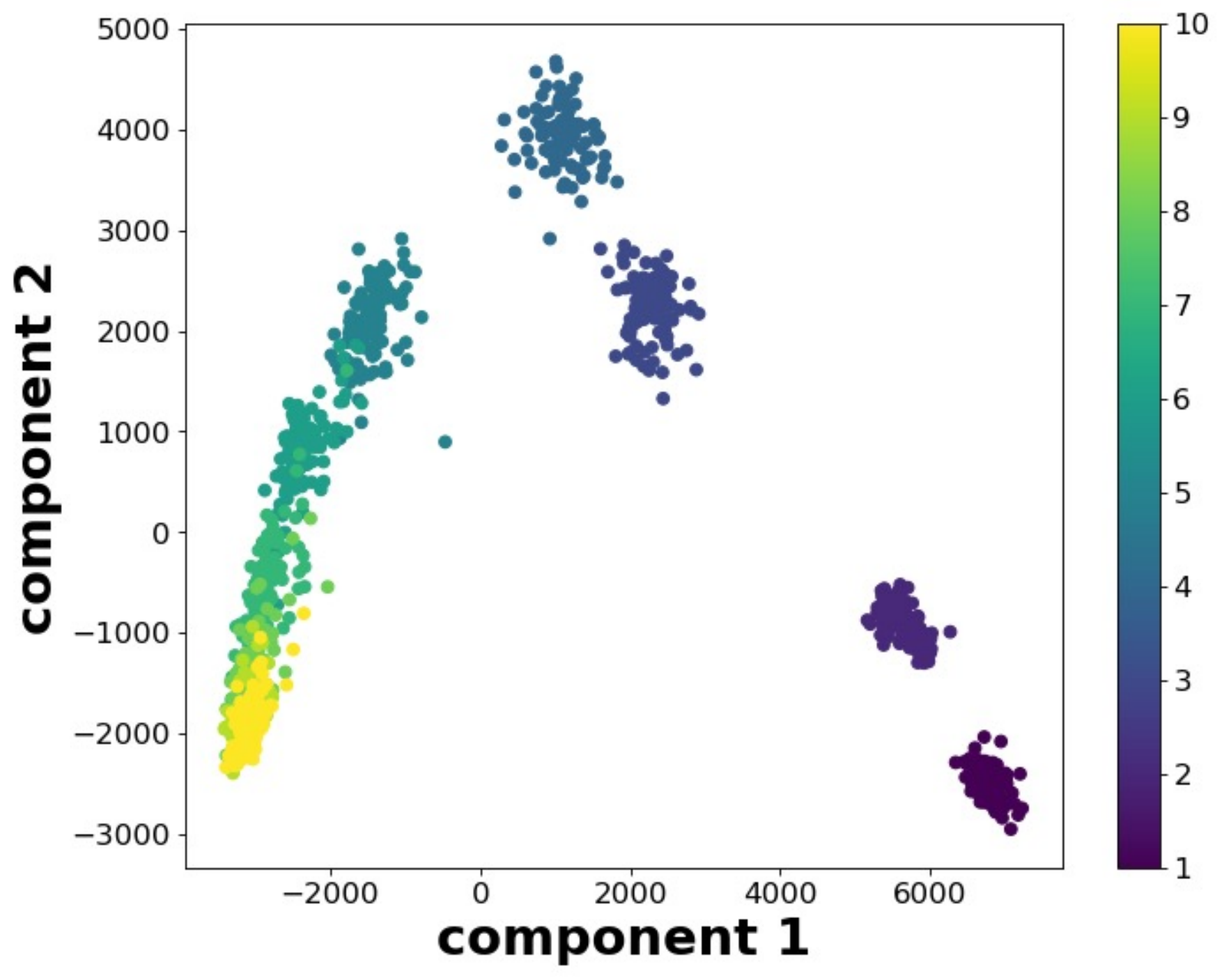} \\
\begin{minipage}{\w}
\centering

{\footnotesize (a)}
\end{minipage} &
\begin{minipage}{\w}
\centering
{\footnotesize (b)} 
\end{minipage} &

\begin{minipage}{\w}
\centering
{\footnotesize (c)} 
\end{minipage}
\end{tabular}
\vspace{-0.12in}
\caption{\textbf{(a) and (b)}: Average output of the model at different $\Delta snr$ and $\Delta sdr$. The vertical lines at each point in the plot shows 95\% C.I.   \textbf{(c)}: PCA visualization of embeddings capturing audio quality information. \vspace{-1\baselineskip}}
\label{fig:simple_timit_esc}
\end{figure}

The performance on the Quantitative tasks are shown in Fig~\ref{fig:simple_timit_esc}(a) and (b) for $\Delta snr$ and $\Delta sdr$ respectively. On an average the model predicts close to the ground truth,  $\Delta snr$ / $\Delta sdr$, for almost all cases. The standard deviations at most $\Delta snr$ / $\Delta sdr$ levels are small as well.

\looseness=-1
\smallsec{In-variance to Language and Speaker's Gender}
We next evaluate the in-variance and robustness of~\frameworkname\, to certain characteristics of speech content such as language and speaker's gender. We find that our model is fairly robust to unseen languages. Moreover, for a given test recording, it does not matter whether the language or the speaker's gender in the reference is same or not. This supports our key hypothesis that non-matching references are sufficient for quality assessment. Please refer to supplementary material for details. 

\smallsec{Commutativity and Indiscernibility of Identicals}
 We empirically study how well our framework supports these two desirable properties.~\textbf{(1)} \textit{Commutativity} - Quality assessment should stay same for $\mathcal{N}(x_1, x_2)$ and $\mathcal{N}(x_2, x_1)$. Changing the order of inputs should not change the \frameworkname\, score. We find that for only a small fraction (less than \emph{2.5\%}) of the test pairs, changing the order also changes the \frameworkname\, score by more than 2dB. Preference task outputs remains consistent as well (flips on changing the order) for more than \emph{97\%} of the test pairs. \textbf{(2)} \textit{Indiscernibility of Identicals} - When both inputs to the models are same, $\mathcal{N}(x, x)$, the probability outputs of the preference task are close to \emph{0.5} for most of the test cases. This is expected as the model is not able to confidently identify which input is cleaner. The~\frameworkname\, scores are also small in most cases, showing that the model supports this property to a considerable extent. Note that our learning mechanism does not explicitly enforce the framework to have these two properties. Hence, small errors are expected. 

\smallsec{Quality based Retrieval and Visualizations}
In this evaluation, we try to understand the representations learned by the network. More specifically, we consider the outputs of the temporal learning block as the Quality Embeddings (QE) and use it for visualizations and quality based retrieval. 
We first create a dataset of 1000 recordings at 10 different discrete quality levels (100 per quality level). In the quality based retrieval task, the goal is to retrieve recordings of same quality level for a given query speech recording. We argue that the QEs can be used to represent audio recordings for such a quality based retrieval. We use Precision@K to assess the top-$K$ retrievals. The mean precision@K over all test queries turns out to be $\emph{0.97}$ for $K=10$ and $\emph{0.95}$ for $K=25$. These high precision retrievals show that the embeddings indeed capture quality.   

Further, Fig~\ref{fig:simple_timit_esc}(c) shows a PCA visualization of the embeddings. We see that the separation between the clusters increases as the quality goes down. For higher qualities, the inter-cluster distances are smaller,  although the embeddings at the same quality level are still tightly clustered together. Moreover, we also observe the cluster centers lie on approximately two piece-wise linear functions, one for low quality (1 to 4) and another for high quality (5 to 10). In other words, the model has implicitly learned high and low quality functions. 

\vspace{-0.15in}
\subsection{Subjective Evaluations}
\label{sec:subjective_validation}
\vspace{-0.1in}
\begin{table}[t!]

\vspace{-0.25in}
\centering
\resizebox{\columnwidth}{!}{
 \begin{tabular}{l l c c c c c c c c}
 \toprule
  \multirow{2}{*}{\bf Type} & \multirow{2}{*}{\bf Name} & \multicolumn{2}{c}{\bf VoCo~\cite{jin2017voco}} & \multicolumn{2}{c}{\bf Dereverb~\cite{su2019perceptually}}& \multicolumn{2}{c}{\bf HiFi-GAN~\cite{su2020hifi}}& 
 \multicolumn{2}{c}{\bf FFTnet~\cite{jin2018fftnet}}\\
 \cmidrule(lr){3-4} \cmidrule(lr){5-6} \cmidrule(lr){7-8} \cmidrule(lr){9-10}
 & &\bf PC &\bf SC & \bf PC & \bf SC & \bf PC & \bf SC & \bf PC & \bf SC\\
 \cmidrule(lr){1-10}
 
 \multirow{2}{*}{\bf Full-ref.}
 & {\bf PESQ} & 0.68 & 0.43 & \bf 0.86 & 0.85	& 0.72	& 0.7 & 0.51 & 0.49 \\
 & {\bf CDPAM} & - & \bf 0.73 & - & \bf 0.93 & - & 0.68 & - & \bf 0.68\\
 \cdashline{1-10}
 
 \multirow{1}{*}{\bf Non-Int.}
  & {\bf DNSMOS} & 0.6 & 0.48 & 0.7	& 0.73 & \bf 0.93 & \bf 0.88 & \bf 0.59 & 0.53 \\

  \cdashline{1-10}
  
 \multirow{4}{*}{\bf \frameworkname\,}
 & {\bf Paired} & 0.64 & 0.6 & 0.46 & 0.65 &  0.59 & 0.81 & 0.46 & 0.47  \\
 
 & {\bf $\text{Unpaired}$} & 0.88$\pm$0.01 &	0.41$\pm$0.06 & 0.63$\pm$0.01	& 0.75 $\pm$0.02 &	0.63$\pm$0.01	& 0.71$\pm$0.01 &	0.46$\pm$0.01 &	0.51$\pm$0.02 \\

   & {\bf \hspace{0.5mm} +$\text{Local-Fixed}$} & \bf 0.89$\pm$0.01 & 0.44$\pm$0.06 & 0.63$\pm$0.01 & 0.75$\pm$0.01 & 0.61$\pm$0.01 & 0.73$\pm$0.01 & 0.46$\pm$0.01 & 0.51$\pm$0.02 \\
 
 & {\bf \hspace{0.5mm} +$\text{Global-Fixed}$} & 0.85$\pm$0.01 & 0.68$\pm$0.03 & 0.66$\pm$0.02 & 0.67$\pm$0.02 & 0.68$\pm$0.01 & 0.78$\pm$0.01 & 0.33$\pm$0.01 & 0.44$\pm$0.01\\


 \bottomrule
\end{tabular}
}
\vspace{-0.1in}
\caption{\textbf{MOS Correlations (1)}: for \frameworkname\,, DNSMOS, PESQ and CDPAM. Spearman (SC), Pearson (PC) correlations are shown. All unpaired \frameworkname\, are obtained using $n=100$ NMRs. $\uparrow$ is better.}
\label{tab:mos_1}
\end{table}

\begin{table}[t!]
\vspace{0\baselineskip}
\centering
\resizebox{\columnwidth}{!}{
 \begin{tabular}{l l c c c c c c c c}
 \toprule
  \multirow{2}{*}{\bf Type} & \multirow{2}{*}{\bf Name} &  
 \multicolumn{2}{c}{\bf PEASS~\cite{emiya2011subjective}} & 
 \multicolumn{2}{c}{\bf VCC-2018~\cite{lorenzo2018voice}} & 
 \multicolumn{2}{c}{\bf Noizeus~\cite{hu2007subjective}} &
 \multicolumn{2}{c}{\bf TCD-VoIP~\cite{harte2015tcd}}\\
 \cmidrule(lr){3-4} \cmidrule(lr){5-6} \cmidrule(lr){7-8} \cmidrule(lr){9-10} 
 & &\bf PC &\bf SC & \bf PC & \bf SC & \bf PC & \bf SC & \bf PC & \bf SC  \\
 \cmidrule(lr){1-10}

   \multirow{2}{*}{\bf Full-ref.}
 & {\bf PESQ} &  \bf 0.86 & 0.71 & \bf 0.51 & 0.56 & 0.43 & 0.42 & \bf 0.89 & \bf 0.90 \\
 & {\bf CDPAM} &  - & \bf 0.74 & - & \bf 0.61 & - & \bf 0.71 & - & 0.88 \\
 \cdashline{1-10}
 \multirow{1}{*}{\bf Non-Int.}
  & {\bf DNSMOS} &  0.39 & 0.21 & 0.37 & 0.42 & 0.41 & 0.59 & 0.71 & 0.72 \\
    \cdashline{1-10}

 \multirow{4}{*}{\bf \frameworkname\,}
 & {\bf Paired} & 0.26 & 0.43 & 0.48 & 0.39 & 0.47 & 0.46 & 0.38 & 0.44 \\
 
 & {\bf $\text{Unpaired}$} & 0.38$\pm$0.01 & 0.40$\pm$0.01 & 0.61$\pm$0.01 & 	0.41$\pm$0.02  &	\bf 0.50$\pm$0.02 & 	0.39$\pm$0.05 & 0.43$\pm$0.01	& 0.46$\pm$0.02 \\

   & {\bf \hspace{0.5mm} +$\text{Local-Fixed}$} &  0.40$\pm$0.04 & 0.52$\pm$0.06 & 0.65$\pm$0.04 & 0.39$\pm$0.02 & 0.45$\pm$0.01 & 0.44$\pm$0.02 & 0.43$\pm$0.02 & 0.41$\pm$0.04\\

 & {\bf \hspace{0.5mm} +$\text{Global-Fixed}$} &  0.41$\pm$0.05 & 0.57$\pm$0.05 & 0.47$\pm$0.01 & 0.41$\pm$0.01 & 0.48$\pm$0.02 & 0.51$\pm$0.01 & 0.56$\pm$0.01 & 0.52$\pm$0.03\\
 \bottomrule
\end{tabular}
}
\vspace{-0.1in}
\caption{\textbf{MOS Correlations (2)}: for \frameworkname\,, DNSMOS, PESQ and CDPAM. Spearman (SC), Pearson (PC) correlations are shown. All unpaired \frameworkname\, are obtained using $n=100$ NMRs. $\uparrow$ is better. \vspace{-1\baselineskip}}
\label{tab:mos_2}

\end{table}

\looseness=-1
The subjective evaluations are aimed at assessing~\frameworkname\, as a proxy for subjective judgments by humans. More specifically, we try to understand how well~\frameworkname\, correlates with MOS across different speech tasks. We also examine~\frameworkname's suitability in subjective 2-alternative forced-choice (2AFC) tests~\cite{manocha2020differentiable}. 

We consider an exhaustive set of 10 different datasets for this evaluation. These datasets span over a variety of well-known speech problems; \textbf{(1)} Speech Synthesis (VoCo~\cite{jin2017voco} and FFTnet~\cite{jin2018fftnet}), \textbf{(2)} Speech Enhancement (Dereverberation~\cite{su2019perceptually}, Noizeus~\cite{hu2007subjective}, HiFi-GAN~\cite{su2020hifi}), \textbf{(3)} Voice Conversion (VCC-2018~\cite{lorenzo2018voice}), \textbf{(4)} Speech Source Separation (PEASS~\cite{emiya2011subjective}), \textbf{(5)} Telephony Degradations~\cite{harte2015tcd}, \textbf{(6)} Bandwidth Extension (BWE~\cite{feng2019learning}), and \textbf{(7)} General Degradation's (Simulated~\cite{manocha2020differentiable}). Please refer to supplementary material for details about these datasets. 


\smallsec{MOS Correlations}
All of the above datasets come with MOS ratings for each audio recording. We evaluate our framework by computing Pearson Correlation Coefficient (PC) and Spearman's Rank Order Correlation (SC) of \frameworkname\, scores with MOS ratings on each dataset. Since MOS scores are always obtained keeping a clean recording in mind, we obtain ``absolute" \frameworkname\, scores (Sec~\ref{sec:usage}) by using clean references. As mentioned in Sec~\ref{sec:usage}, we can get a more reliable score by averaging over multiple references. The results discussed in this section are for $n = 100$ and an ablation over $n$ are provided in further sections. 

Finally, to decouple the effect of source/type of references on the results, we compute \frameworkname\, for a given test recording in 4 different ways. \textbf{(i)} \emph{Paired}: This is a special case where matched clean recordings are given as references to \frameworkname. Obviously $n=1$ in this case. \textbf{(ii)} \emph{Unpaired}: In this case, for the given test recording, a clean NMR is randomly selected from the same dataset. The process is repeated $n=100$ times and the scores are average to obtain the final \frameworkname\, for the test recording.  \textbf{(iii)} \emph{Unpaired Local-Fixed}: the same set of NMRs are used for all test recordings. The set ($n=100$) is fixed locally for each dataset. \textbf{(iv)} \emph{Unpaired Global-Fixed}: this is similar to the previous case, where the same set of NMRs are used for all test recordings. However, in this case, this set is fixed across all datasets. The NMRs ($n=100$) are selected from the DAPS dataset~\cite{mysore2014can} and used for evaluations on all 8 datasets. All unpaired experiments are repeated 10 times and average results with standard deviations are reported. 

\looseness=-1
Tables~\ref{tab:mos_1} and \ref{tab:mos_2} shows correlations with MOS on all 8 datasets. Correlations for full-reference SQA methods (PESQ and CDPAM), and non-intrusive DNSMOS are also shown. First, we note that \frameworkname\,, is not only competitive to these methods but it even surpasses their performance in several cases. Our method is better than DNSMOS on 4 datasets  (VoCo, PEASS, VCC and Noizeus) and is very competitive on others. Note that, DNSMOS is \emph{explicitly} trained on a large scale MOS dataset. Our method is not trained on any perceptual labels or judgments and is still almost as good as DNSMOS. The results from the tables also show that our method is a good substitute for full-reference methods in several cases. However, unlike these full-reference methods, our framework stays useful in practical and real-world situations where the clean matching reference might not be available.  

\smallsec{2AFC Tests}
2AFC test is a comparative approach to subjective evaluations. In this case, listeners are given a reference and two test recordings and asked to judge which one sounds more similar (in terms of quality) to the reference. We follow the evaluation protocol of Manocha et al.~\cite{manocha2020differentiable} and report how accurately different methods can predict same judgement as humans. Table~\ref{table_triplet} shows accuracy of different methods on 4 datasets. Full-reference methods give best performances in most cases. However, our framework, \frameworkname\, is substantially better than DNSMOS in all cases. 

\smallsec{Ablations}
To further expound~\frameworkname\,, we conducted ablation studies for two factors. Please refer to the supplementary material for detailed results and analyses of these ablations. 

\textit{Multi-objective Learning of Quantification Task}: In this ablation, we try to assess the significance of SNR and SI-SDR heads for the Quantification task in \frameworkname. Our ablation shows that using both heads is superior to using just SNR or SI-SDR head; often the improvement is more than $30$ to $40\%$ over using just one of the heads. SI-SDR is also performs better in all cases when it comes to having just SNR or SI-SDR for the quantification task.
This is expected, as with SI-SDR, we can use a wide variety of degradations, whereas SNR based training can only use additive noise degradations. 

\textit{Number of NMRs ($n$)}: In this ablation, we examine how the MOS correlations are affected by the number of NMRs used for each test recording. Increasing the number of NMRs for \frameworkname\, improves correlations with MOS. Depending on the dataset, this improvement can be as much as  $10$ to $15$\% when $n$ is increased from $1$ to $100$. 

\begin{table}[t!!]
\vspace{-0.2in}
\RawFloats
    \begin{minipage}{.50\linewidth}
    \vspace{-0.05in}
    \centering
    \begingroup
    \renewcommand{\arraystretch}{1.88}
      \setlength{\tabcolsep}{7pt}
      \resizebox{\columnwidth}{!}{
        \begin{tabular}{l c c c c}
             \toprule
             {\bf Name}  & {\bf Simulated~\cite{manocha2020differentiable}}& {\bf FFTnet~\cite{jin2018fftnet}}& {\bf BWE~\cite{feng2019learning}} & {\bf HiFi-GAN~\cite{su2020hifi}} \\
             \toprule
             {\bf PESQ} & 86.0 & 67.0 & 38.0 & 88.5 \\
              {\bf CDPAM} & \bf 87.7 & \bf 88.5 & \bf 75.9 & \bf 96.5 \\
              \cdashline{1-5}
              {\bf DNSMOS} & 49.2 & 58.8 & 45.0 & 62.3 \\
             \cdashline{1-5}
              {\bf \frameworkname\,} & 68.7 & 73.3 & 53.3 & 81.6 \\
             \bottomrule
            
            \end{tabular}}
            \caption{Accuracy on 2AFC predictions }
            \label{table_triplet}
            \endgroup
    \end{minipage}%
    \hspace{2mm}
    \begin{minipage}{.50\linewidth}
      \centering
      \setlength{\tabcolsep}{3pt}
      \resizebox{\columnwidth}{!}{
            \begin{tabular}{l l c c c c c c}
             \toprule
             {\bf Type} & {\bf Data\%} &  {\bf PESQ} & {\bf STOI} & {\bf SNRseg} & {\bf CSIG} & {\bf CBAK} & {\bf COVL}
              \\ \toprule
              {\bf Noisy} &  
             &  1.97 & 91.50 & 1.72 & 3.35 & 2.44 & 2.63 \\
             \cdashline{1-8}
             \multirow{3}{*}{\bf Baseline} & 33\% 
             & 2.22  & 91.7  &  8.18 & 3.26 & 2.98 & 2.72 \\
             & 66\% 
             & 2.30 & 92.23 & 8.54 & 3.45 & 3.04 & 2.85 \\
             & 100\%
             & 2.39 & 91.89 & 8.71 & 3.55 & 3.10 & 2.95 \\
             \cdashline{1-8}
             \multirow{3}{*}{\bf Pre-trained}
             & 33\%
             & 2.28 & 92.30 & 8.33 & 3.43 & 3.03 & 2.83 \\
              & 66\% 
             & 2.35 & 92.90 & 8.77 & 3.53 & 3.1 & 2.92 \\
             & 100\% 
             & \bf 2.46 & \bf 93.53 & \bf 8.81 & \bf 3.59 & \bf 3.17 & \bf 2.99 \\
             \bottomrule
            \end{tabular}}
            \vspace{-0.1in}
            \caption{Evaluation of~\frameworkname\, pre-training for speech enhancement.}
            \label{table_denoising}
    \end{minipage} 
\vspace{-2.0\baselineskip}
\end{table}

\subsection{Speech enhancement}
\label{sec:se}
In this section, we show the utility of \frameworkname\ for training a Speech Enhancement (SE) system. Our SE network is based on a popular U-Net type convolutional recurrent neural network~\cite{tan2019learning}. The baseline SE model is trained in a supervised manner using matched clean and noisy speech pairs.

Note that the idea behind using \frameworkname\ for SE is quite different from various works that use a speech enhancement model to guide training~\cite{fu2019metricgan}. These methods rely on metrics to compute a loss function (generally in addition to L1/L2 losses w.r.t the target clean speech). They train local differentiable proxies of quality (e.g., PESQ) or intelligibility (e.g., STOI) at every iteration of training. Clearly, this requires matching noisy-clean pairs for training. MetricGAN does this in a GAN framework for improving the discriminator. Our approach for using \frameworkname\ for SE is very different. Since \frameworkname\ can compare two non-matched recordings, we use it to pre-train the Speech Enhancement model in a way that does not require the exact matched pairs of recordings. This can be much more useful for out-of-domain data adaptation, unseen noise conditions, and other sparse labeled-data situations. Our approach can leverage potentially large amounts of unpaired data in these cases. Moreover, as already mentioned in Sec~\ref{intro}, using certain metrics such as PESQ as the target quality has its own issues, including sensitivity to perceptually invariant transformations especially in cases where the quality between two recordings to be compared are perceptually close.

\looseness=-1
We show that pre-training the network using \frameworkname\, as loss function can lead to improvements in performance. More specifically, given a noisy speech recording and a non-matching clean speech recording, during pre-training stage, the SE model tries to minimize the \frameworkname\, score between the output of the SE model and clean recording. Since we do not require \textit{any} matched clean and noisy pairs for this step, we can leverage unlimited amount of training data - most significantly noisy recordings from real world for which the corresponding clean recordings are not available. After the pre-training stage, the model can be fine-tuned for enhancement following the usual paired training strategy. We use the VCTK dataset~\cite{valentini2017noisy} for these experiments and consider 3 baseline SE models. These are models trained on $1/3^{rd}$, $2/3^{rd}$ and full training set. Following prior works on this dataset, we report enhancement performance on 5 different objective metrics. 

Table~\ref{table_denoising} shows the benefits of \frameworkname\, based pre-training. We see that pre-training consistently improves all 5 metrics for all 3 baseline models. We further analyse the improvements at different SNR levels, and observe that \frameworkname\, based pre-training is most effective in high SNR conditions, where the degradation caused by strong background noise is not noticeable and harder to learn. Please refer to the supplementary material for further details. 
Our motivation behind these experiments is to demonstrate proof of concept of \frameworkname\ in an application (e.g., SE). Hence, our focus is not on obtaining SOTA performance on the VCTK dataset. The improvements coming from pre-training with \frameworkname\ show the potential of leveraging the differentiable \frameworkname\ in different downstream tasks.

\section{Conclusion}
\label{sec:conclusion}
In this paper, we presented a new framework for speech quality assessment. Motivated by human's ability to assess quality independent of the speech content, we propose a framework for SQA using non-matching references. This can potentially open up a new research direction in the development of SQA methods. In subjective evaluations, our method works as competently as established full-reference and subjective non-intrusive SQA methods. At the same time, it addresses key limitations of those methods. Going forward, our focus will be on developing novel methods under the \frameworkname\, framework which can correlate better with subjective human ratings.

\bibliographystyle{unsrtnat}
\bibliography{ref}

\clearpage

{\centering\textbf{\huge Appendix}}

\appendix


\section{NORESQA Architecture}

\frameworkname's architecture comprises of three key components: a \emph{feature-extraction} block, a \emph{temporal-learning} block and \emph{task-specific} heads.

\smallsec{Feature-Extraction Block}

 We use the Inception architecture in the feature extraction block. 
 It consists of 4-block Inception modules: each module consisting of 64 convolutional filters - \textit{24} 1x1 filters, \textit{32} 3x3 filters, and \textit{8} 5x5 filters. These filters are concatenated and finally passed through a 1x4 maxpool block, preserving the temporal dimension. This is repeated for each of the 4 inception blocks. The final output dimensions of the model is $B \times 64 \times T \times 2$, where $B$ is the batch size, and $T$ is the number of frames. ReLU is used as the activation function after each layer.
 
\smallsec{Temporal Learning Block}

We use temporal convolutional networks (TCNs) in the temporal learning block.
The network consists of 4 temporal blocks. Each block consists of 2 convolutional layers with a kernel size of 1x3, with each layer consisting of 32, 64, 64 and 128 channels for each of the 4 blocks respectively. After the convolutional layers, each layer uses weight normalization which is a reparameterization trick that decouples the magnitude of a weight tensor from its direction. The outputs are then passed to ReLU as an activation function, and finally to dropout having a value of $0.2$. The convolutional layers in each block consist of dilated convolutions with dilation factors of 2, 4, 8 and 16 for each of the 4 blocks. Use of dilated convolutions increases the effective history of our model. The initial weights of this network are chosen from the normal distribution $\mathcal{N}(0,1e^{-2})$.


The parameters for both these blocks (feature-extraction and temporal-learning) are shared between the two inputs to our model. 
The embeddings for each input are concatenated (along the channel dimension), and are then passed next to the task-specific heads. The input to this model is $B \times T \times 128$, and the output is also $B \times T \times 128$, since TCNs can maintain the same length of the signal.

\smallsec{Task Specific Heads}

Each of the two tasks (\emph{preference-task} amd \emph{quantification-task}) each has a separate head. Their architecture is described next:

\textit{Preference Task:}
 The output of the preference task is a frame-level prediction of which input is cleaner. This head consists of 3 convolutional layers, each consisting of 32, 8 and 2 channels respectively. The kernel size for each layer is 1x5, with each layer also having BatchNorm and dropout (0.2). The input to this model is $B \times T \times 256$ (after concatenating along the channel dimension of the two inputs) and the output is $B \times T \times 2$ with a framewise prediction of which input is cleaner.

\textit{Quantification Task:}
Refer to Sec 3.3 (main paper). The objective of the quantification task is to quantify the framewise quality difference between the two inputs. Here we formulate this as a classification problem, where we divide the whole range of SNR ($\MakeUppercase{\Delta}snr_{max})$ and SI-SDR ($\MakeUppercase{\Delta}sdr_{max}$) into $K$ equal intervals. The output of this head is a probability distribution over all $K$ classes. Similar to the Preference task, this network also produces frame-level distributions. 
For both objectives (SNR and SI-SDR), we take $K=40$. This head consists of 3 convolutional layers, each consisting of 64, 50 and 40 channels respectively. The kernel size for each layer is 1x5, with each layer also having BatchNorm and dropout (0.2). The input to this model is again $B \times T \times 256$ (after concatenating along the channel dimension of the two inputs) and the outputs are $B \times T \times 40$ for both SI-SDR and SNR for a framewise prediction of relative quality.

\section{Experimental setup}

\smallsec{Dataset}
For the training and validation set, we choose the clean audio recordings from the DNS Challenge. The noise perturbations are sampled from the FSDK50 dataset that consists of over 51k audio samples encompassing around 100 hours of audio manually labelled using 200 classes drawn from the Audioset ontology. As additional examples of distortions, we also use  Clipping distortion, and Frequency masking as very common examples of distortions found in audio processing tasks. All data is divided into 90\% train and 10\% validation so that there is no overlap of train and validation data.

For the test set, we use the TIMIT dataset for the clean recordings. The noise perturbations are sampled from the ESC-50 dataset that consists of 2000 labeled recordings equally balanced between 50 classes (exactly 40 clips per class). We also use Gaussian noise and Mu law compression as two examples of unseen test distortions.

The training and testing pairs are created by adding the same type/category of noise to both recordings, but at two \emph{different} noise levels to two \emph{different} clean recordings. We design a simulation environment where we sample the clean recordings and noisy recordings from their respective datasets, and create the degraded recordings at differences levels of noise. 

We also include reverberations in our audio recordings to make them more realistic and help the model generalize better.  We sample impulse responses from the DNS Challenge dataset and convolve with the noisy recordings before input to the model. We use the ``medium-room'' and ``small-room'' RIRs, where the length and width of the room are sampled from 1 to 30m.

\section{Objective Evaluations}

\subsection{In-variance to language}

\begin{figure}[h!]

\centering
\setlength{\w}{0.50\columnwidth}
\setlength{\tabcolsep}{2pt}
\begin{tabular}{cc}
\includegraphics[width=\w]{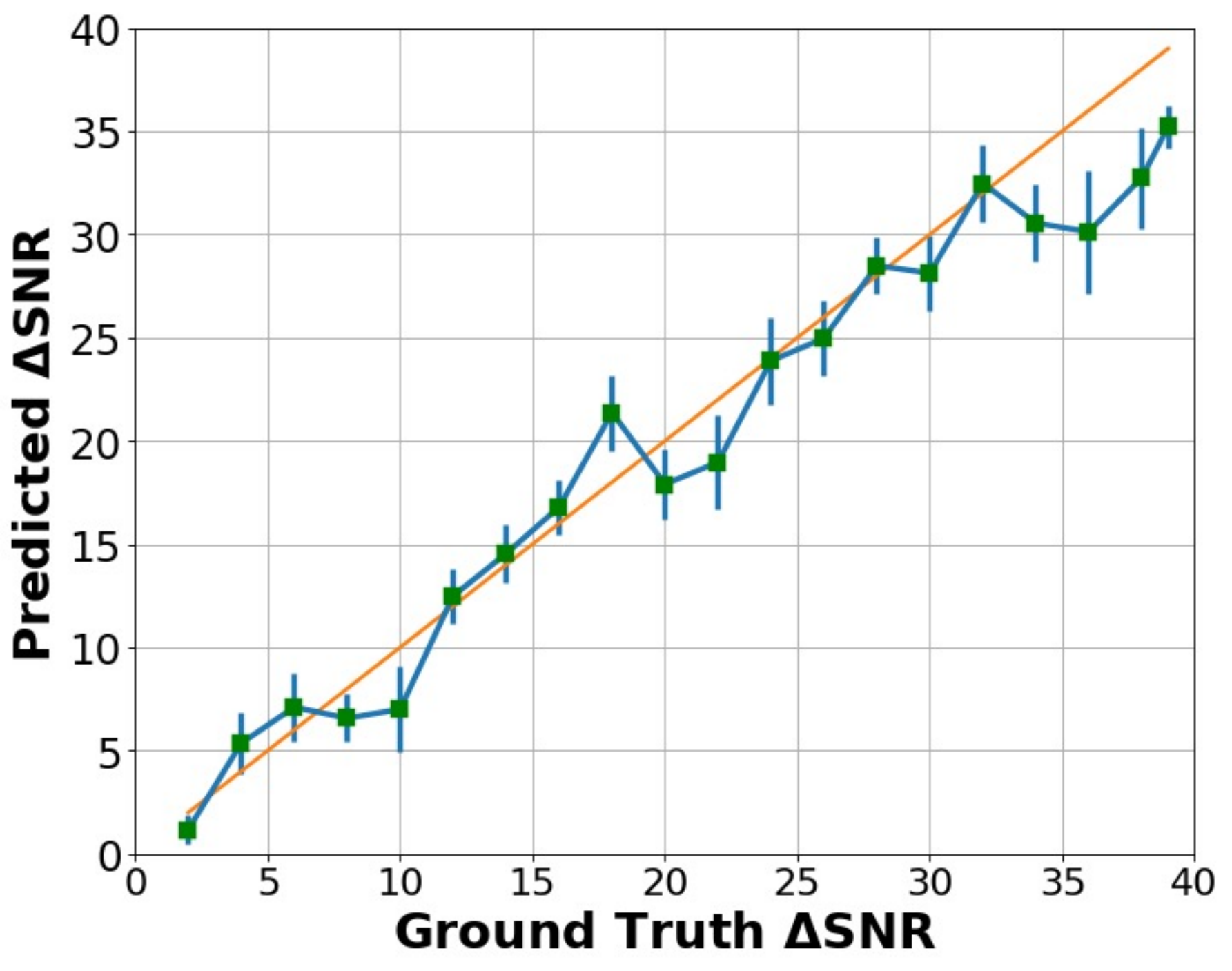} &
\includegraphics[width=\w]{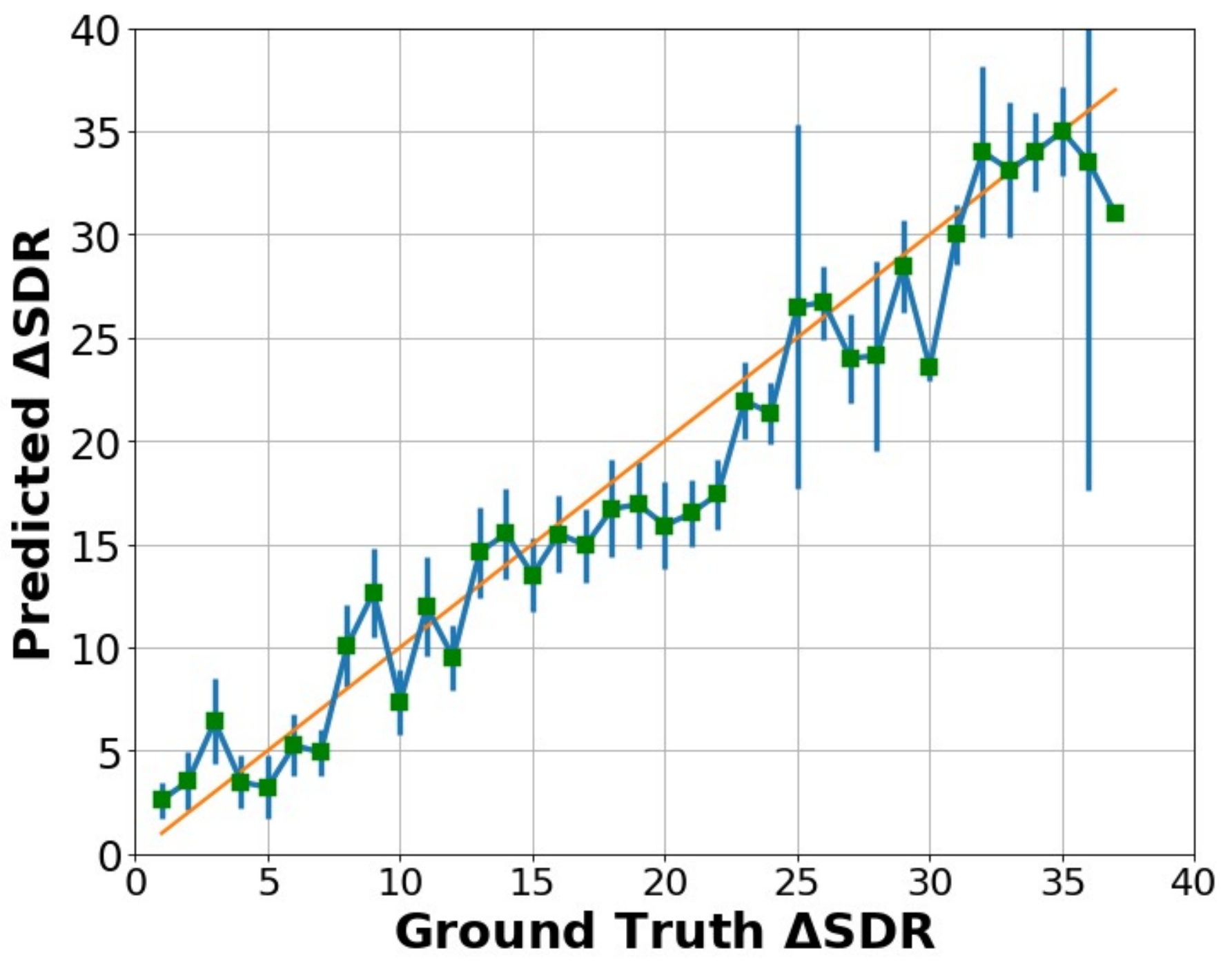} \\
\begin{minipage}{\w}
\centering
{\footnotesize (a)}
\end{minipage} &
\begin{minipage}{\w}
\centering
{\footnotesize (b)} 
\end{minipage}
\end{tabular}
\caption{Variation of our models output with increasing~(a) SNR and~(b) SI-SDR using \emph{test and reference} recordings randomly chosen from Mandarin, French, Arabic,  Turkish, Spanish, Bengali, Gujarati, Hindi and Marathi.}
\label{invariance_english}
\end{figure}

Fig~\ref{invariance_english} shows our metric's outputs with increasing SNR and SI-SDR difference for an unseen set of clean recordings that were randomly chosen from various languages including Mandarin, French - Arabic - Turkish - Spanish, Indian languages - Bengali, Gujarati, Hindi and Marathi, and noise recordings from ESC-50. Note that, the model itself was trained on English speech and we are testing with different languages. We see that the model performs fairly well for these unseen languages.  

\begin{figure}[h!]

\centering
\setlength{\w}{0.50\columnwidth}
\setlength{\tabcolsep}{2pt}
\begin{tabular}{cc}
\includegraphics[width=\w]{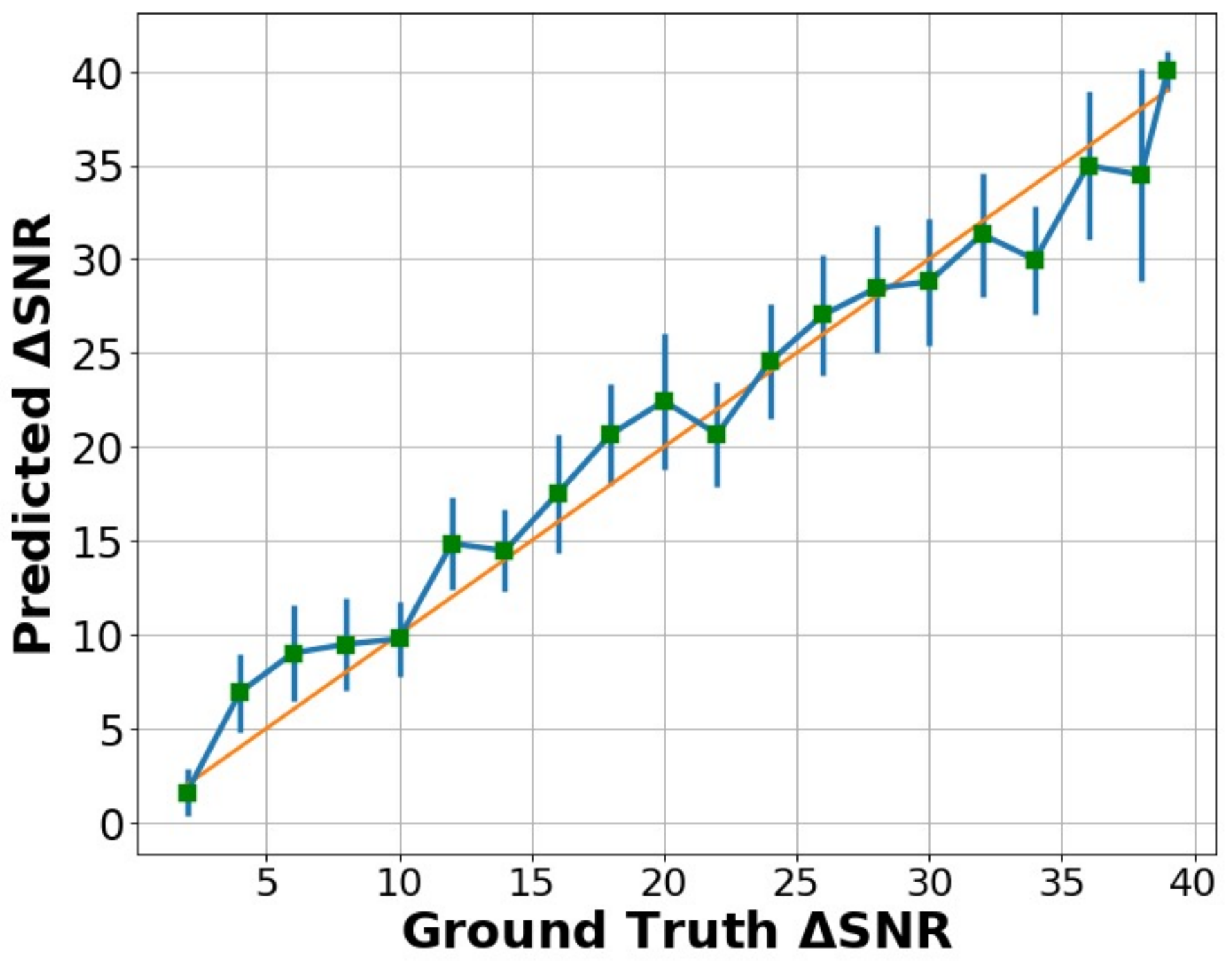} &
\includegraphics[width=\w]{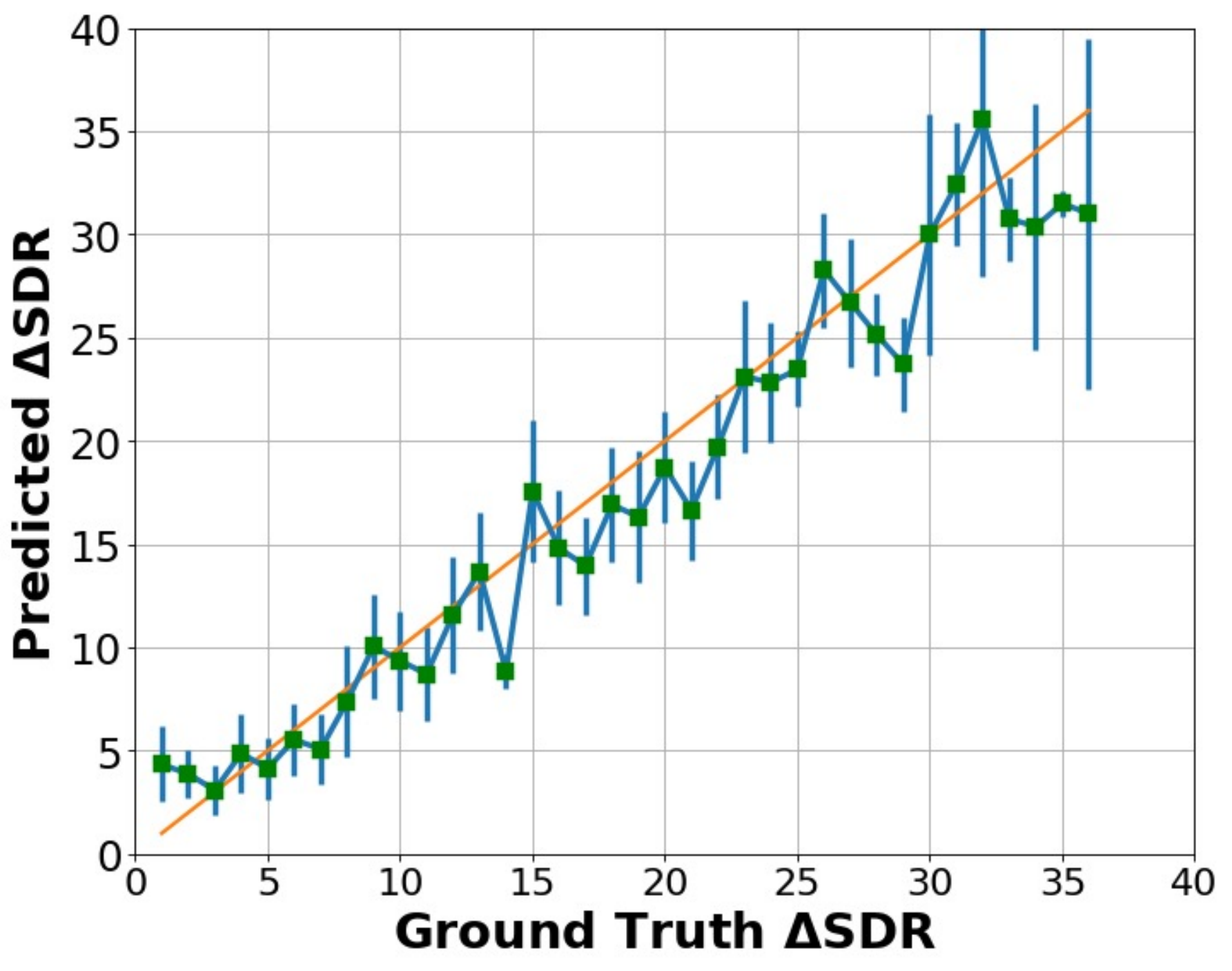} \\
\begin{minipage}{\w}
\centering
{\footnotesize (a)}
\end{minipage} &
\begin{minipage}{\w}
\centering
{\footnotesize (b)} 
\end{minipage}
\end{tabular}
\caption{Variation of our models output with increasing~(a) SNR and~(b) SI-SDR using \emph{test} speech from English (TIMIT), and \emph{reference} recordings from Mandarin, French, Arabic,  Turkish, Spanish, Bengali, Gujarati, Hindi and Marathi.}

\label{invariance_english_ref}
\end{figure}

We studied robustness with respect to language by using non-matching references as well. In this case our test recording is always randomly selected from the English TIMIT dataset, and our reference recording is randomly chosen from any of the multi-langauge datasets (Fig~\ref{invariance_english_ref}).
We observe that the general trend of the model is quite similar to Fig 2(a) and (b), (from the main paper, that showed the variation with English) which suggests that the model is invariant to language, and only considers acoustic differences to assess relative quality. Overall, these evaluations show that the trained model is not only robust to the language of test recording but can also compare quality of two speech signals with different languages.

\begin{figure}[h!]
\centering
\setlength{\w}{0.50\columnwidth}
\setlength{\tabcolsep}{2pt}
\begin{tabular}{cc}
\includegraphics[width=\w]{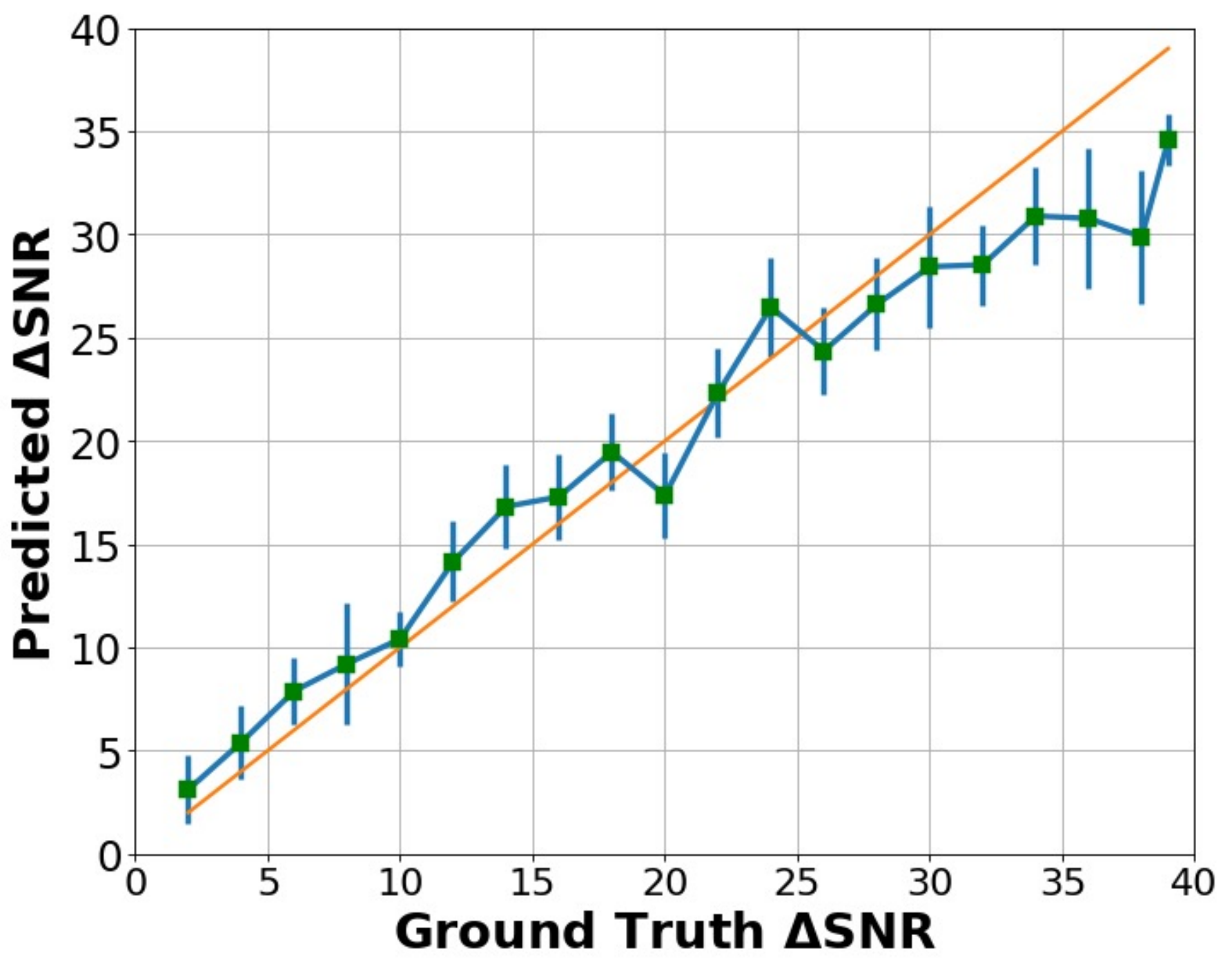} &
\includegraphics[width=\w]{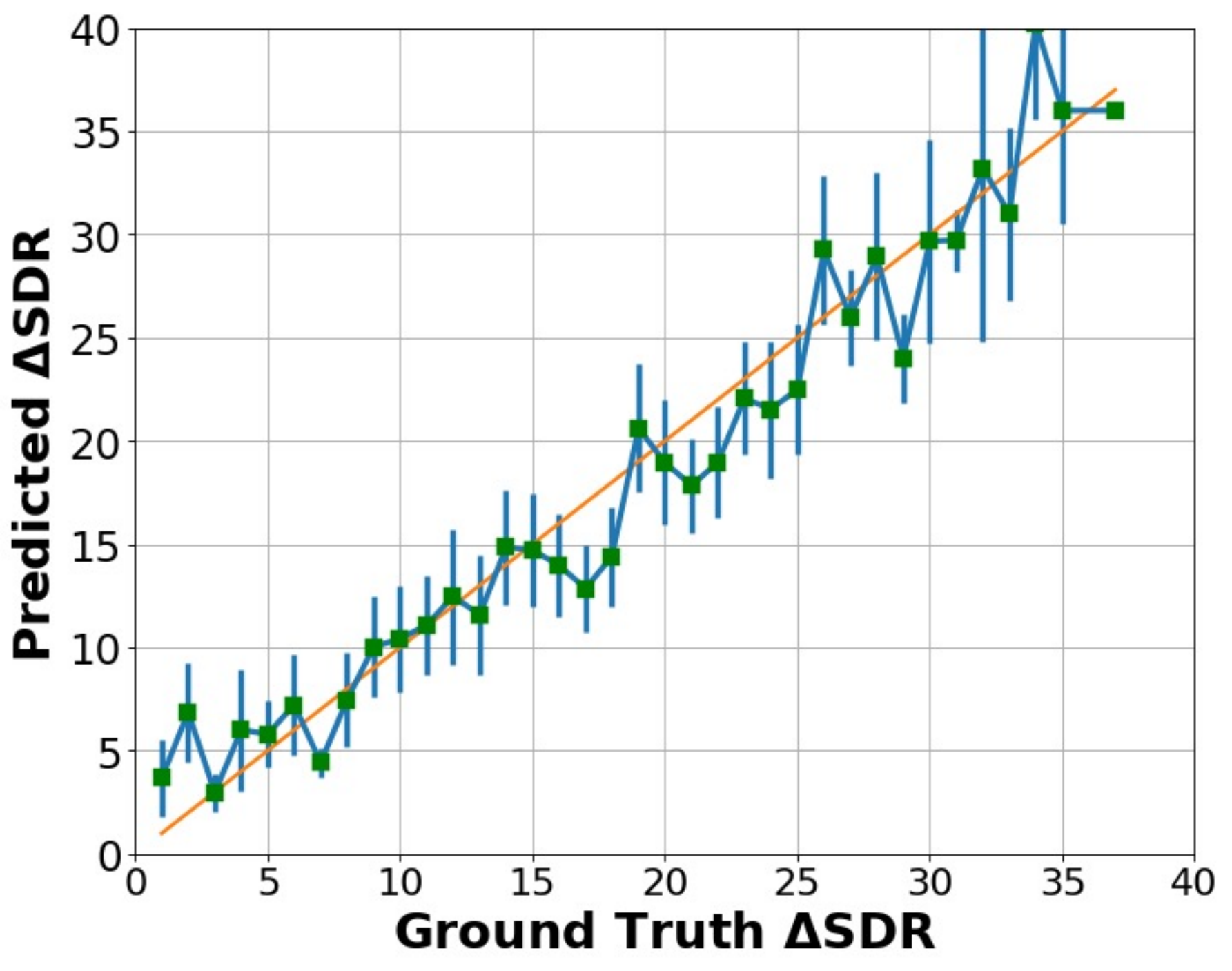} \\
\begin{minipage}{\w}
\centering
{\footnotesize (a)}
\end{minipage} &
\begin{minipage}{\w}
\centering
{\footnotesize (b)} 
\end{minipage}
\end{tabular}
\caption{Variation of our models output with increasing~(a) SNR and~(b) SI-SDR using unseen clean \emph{male} speech from the DAPS dataset.}
\label{invariance_gender_male}
\end{figure}

\begin{figure}[h!]

\centering
\setlength{\w}{0.50\columnwidth}
\setlength{\tabcolsep}{2pt}
\begin{tabular}{cc}
\includegraphics[width=\w]{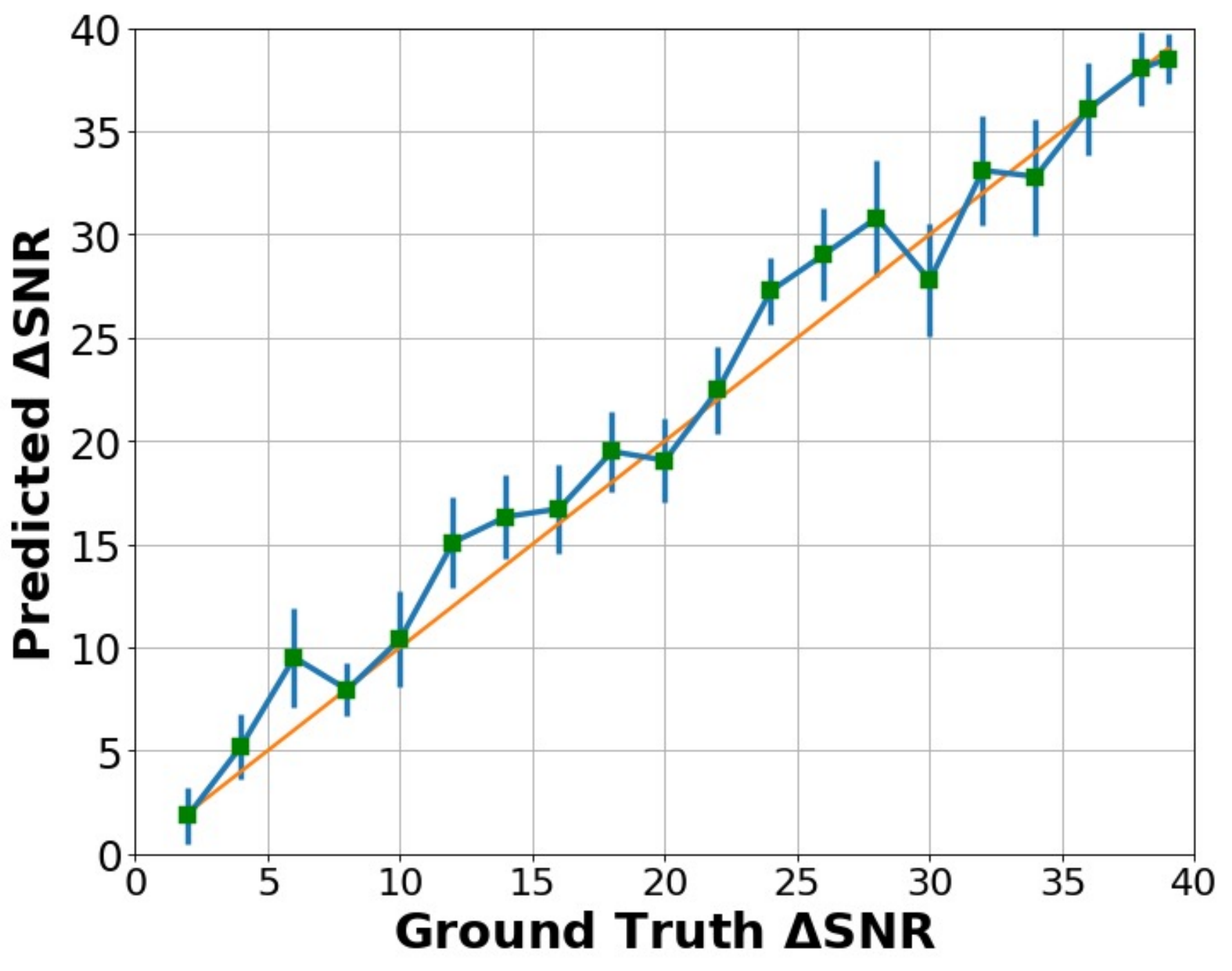} &
\includegraphics[width=\w]{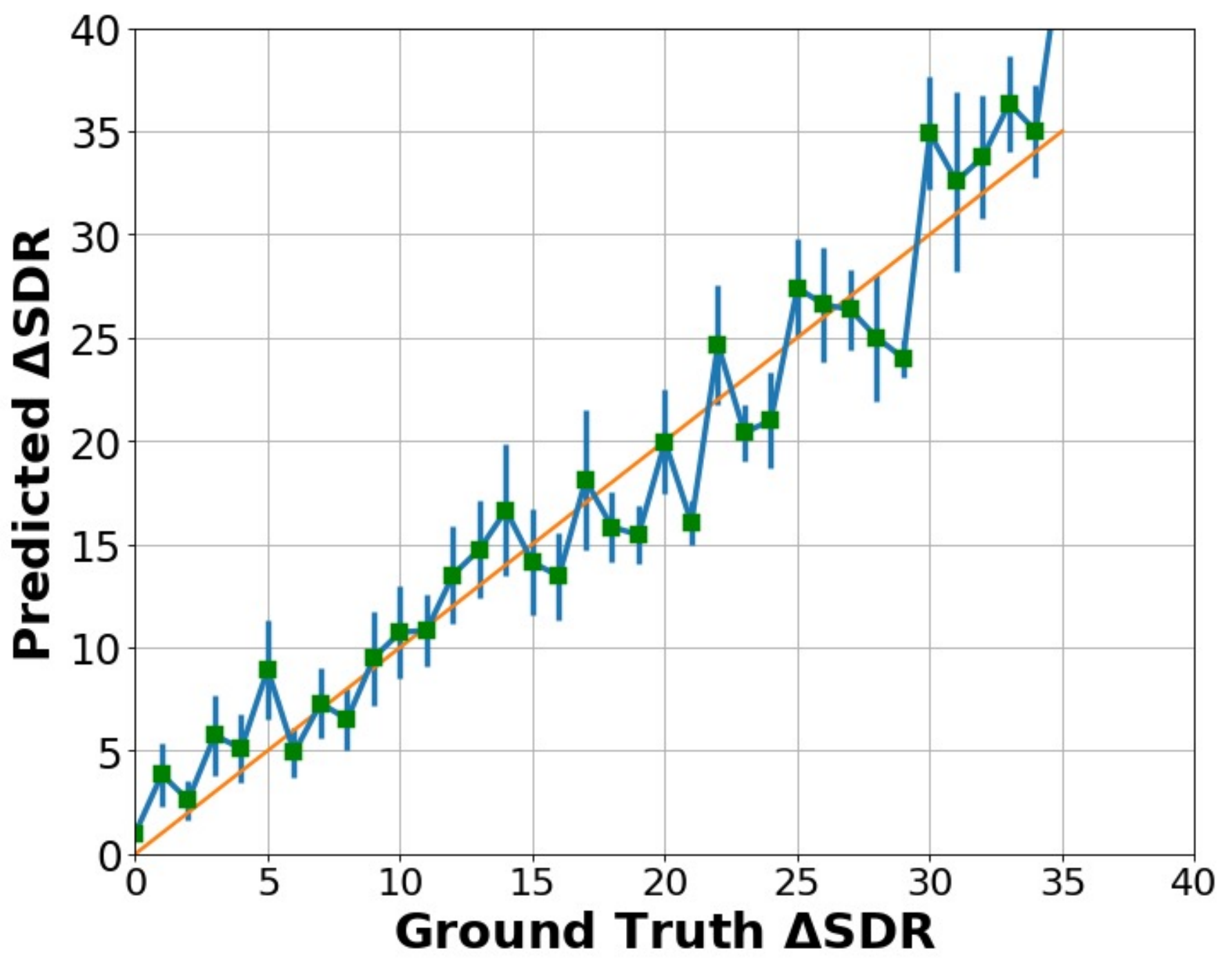} \\
\begin{minipage}{\w}
\centering

{\footnotesize (a)}
\end{minipage} &
\begin{minipage}{\w}
\centering
{\footnotesize (b)} 
\end{minipage}
\end{tabular}
\caption{Variation of our models output with increasing~(a) SNR and~(b) SI-SDR using clean \emph{female} speech from the DAPS dataset.}
\label{invariance_gender_female}
\end{figure}

\begin{figure}[h!]
\centering
\setlength{\w}{0.50\columnwidth}
\setlength{\tabcolsep}{2pt}
\begin{tabular}{cc}
\includegraphics[width=\w]{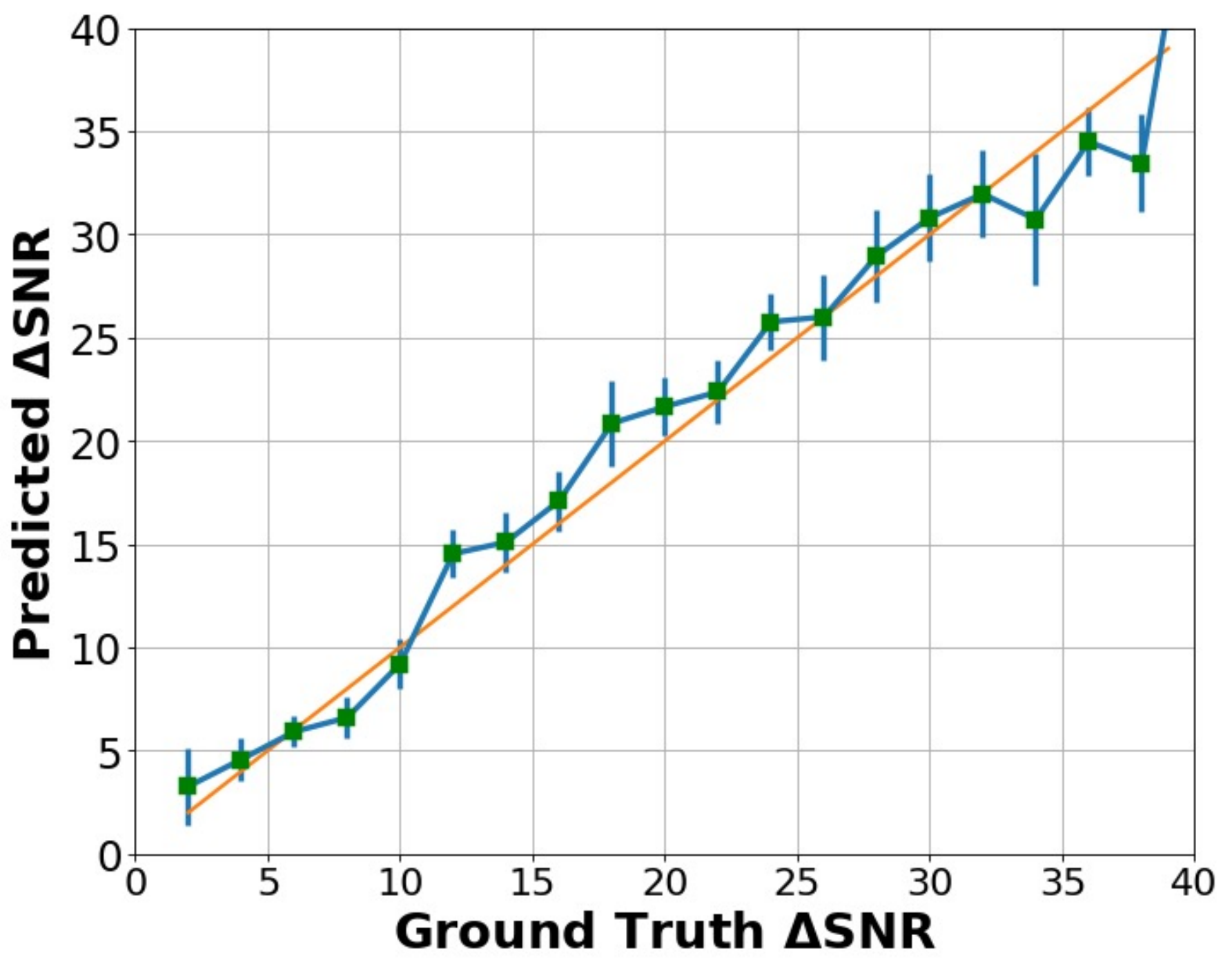} &
\includegraphics[width=\w]{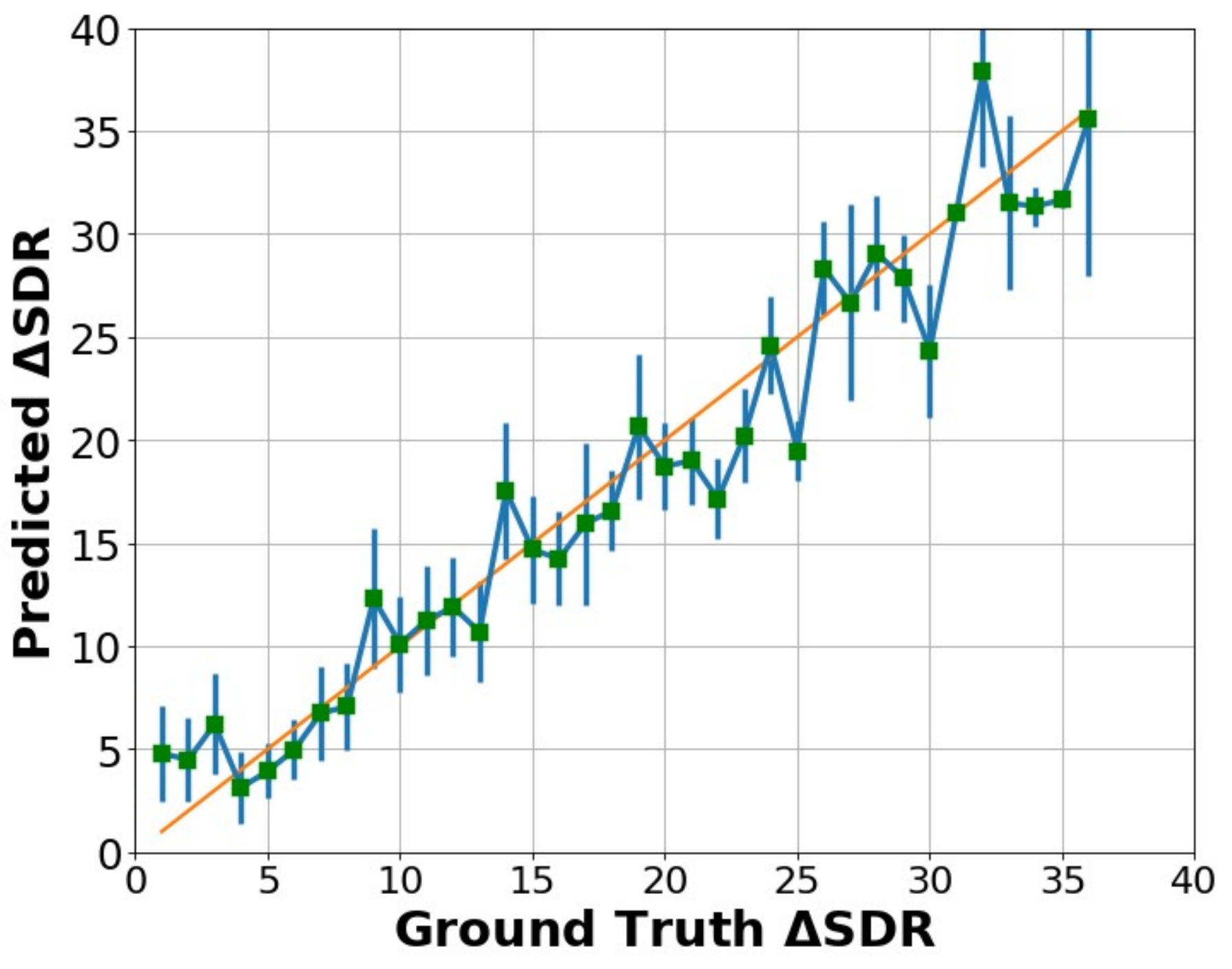} \\
\begin{minipage}{\w}
\centering
{\footnotesize (a)}
\end{minipage} &
\begin{minipage}{\w}
\centering
{\footnotesize (b)} 
\end{minipage}
\end{tabular}
\caption{Variation of our models output with increasing~(a) SNR and~(b) SI-SDR under mismatched gender conditions (i.e., test recording is male speech, and reference recording is female speech and vice-versa)}
\label{invariance_gender_male_ref}
\end{figure}

\subsection{In-variance to gender}
We now evaluate how NORESQA behaves with respect to speaker's gender. Once again we try to disentangle behaviour w.r.t to speaker's gender through two experiments. First, we see how the trained models behave for each gender, male and female. We test the models in male-only speaker condition (test and references are all male speeches) as well as in female-only speaker condition. Fig~\ref{invariance_gender_male} shows the trends for male and~\ref{invariance_gender_female} shows it for female. We note that the model works well in both cases.  

Second, we evaluate how stable NORESQA is when the gender of the test recording \emph{does not} match the reference, i.e., test recording is male speech, and reference recording is female speech and vice-versa (Fig~\ref{invariance_gender_male_ref}). Once again, we observe that mis-matching speaker's gender in test and references does not adversely affect model's behaviour. Overall, based on these evaluation, we conclude that the model is invariant to gender of the speaker and is primarily learning quality related characteristics.

\subsection{Commutativity: $\mathcal{N}(x_{test}, x_{ref})$ = $\mathcal{N}(x_{ref}, x_{test})$}

We empirically evaluate the model to check if it satisfies the commutative property i.e., $\mathcal{N}(x_{test}, x_{ref})$ = $\mathcal{N}(x_{ref}, x_{test})$

To evaluate the \emph{preference-task}, we check the overall recording-wise predictions of both cases (i.e., $\mathcal{N}(x_{test}, x_{ref})$ and $\mathcal{N}(x_{ref}, x_{test})$ and see how well does the quality preference order swap when swapping the order of the inputs. We calculate accuracy by counting the number of times the preference order correctly gets swapped, and divide it by the total number of recordings. Our metric gets an accuracy of \textit{98.3\%} that empirically shows that it obeys the commutative property.

\begin{figure}[h!]
\centering
\includegraphics[width=0.50\columnwidth,]{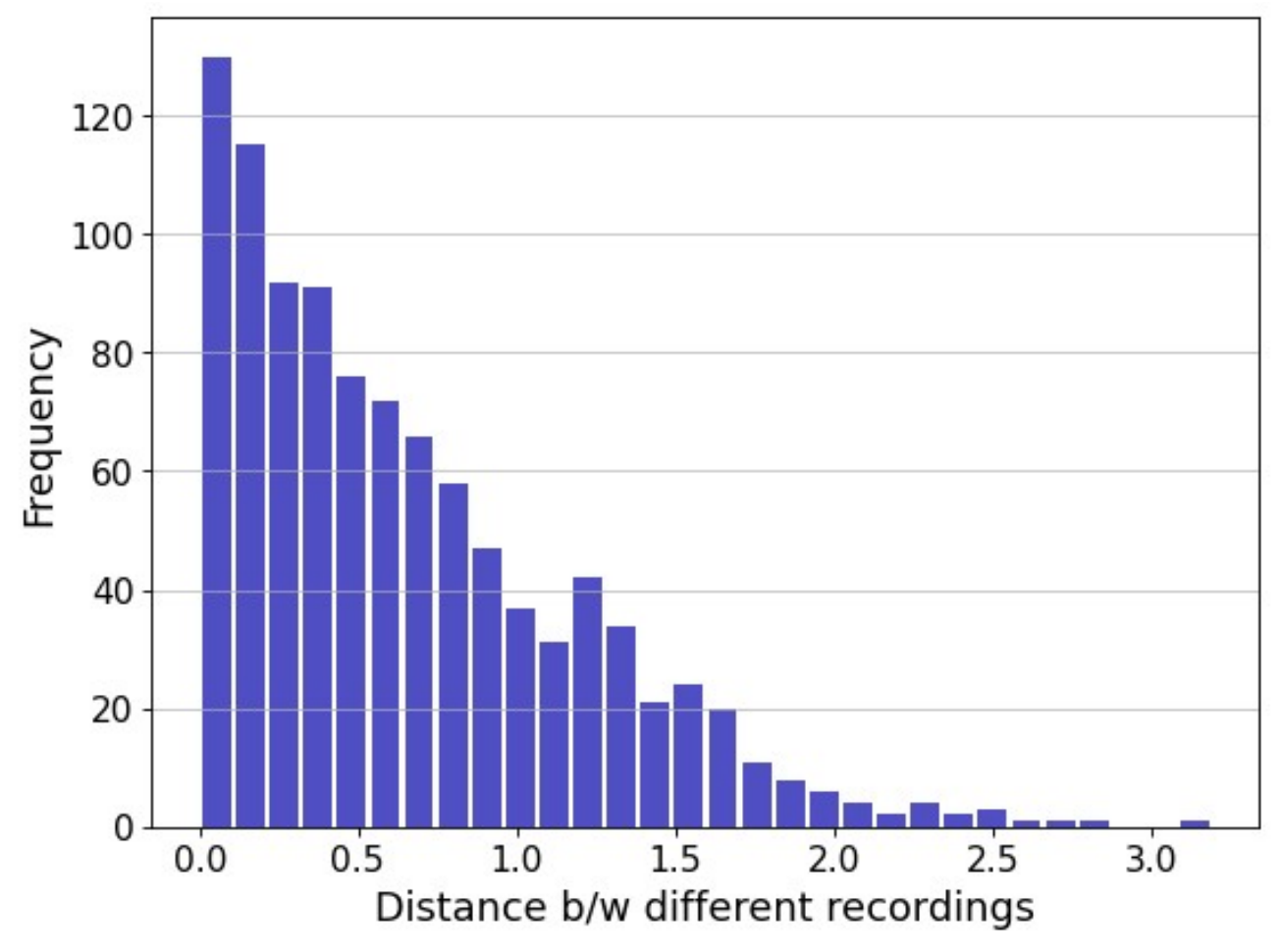}
\caption{{\bf Commutative Property:} Histogram plot of abs($\mathcal{N}(x_{test}, x_{ref})$ - $\mathcal{N}(x_{ref}, x_{test})$)}
\label{f_x_Y}
\end{figure}

To evaluate the \emph{quantification task} (Fig~\ref{f_x_Y}), we compute the predictions of both cases (i.e., $d_1$ = $\mathcal{N}(x_{test}, x_{ref})$ and $d_2$ = $\mathcal{N}(x_{ref}, x_{test})$). We then plot the distribution $abs(d_1-d_2)$ for 1000 different pairs of recordings. We observe that the distribution is centered around 0dB which empirically suggests that it obeys the commutative property.

Overall, our model learns these two desirable properties without any specific training, and this suggests the usefulness of our framework for audio quality judgment.

\subsection{Indiscernibility of Identicals: $\mathcal{N}(x_{test}, x_{test})$ }

\begin{figure}[h!]
\centering
\setlength{\w}{0.50\columnwidth}
\setlength{\tabcolsep}{2pt}
\begin{tabular}{cc}
\includegraphics[width=\w]{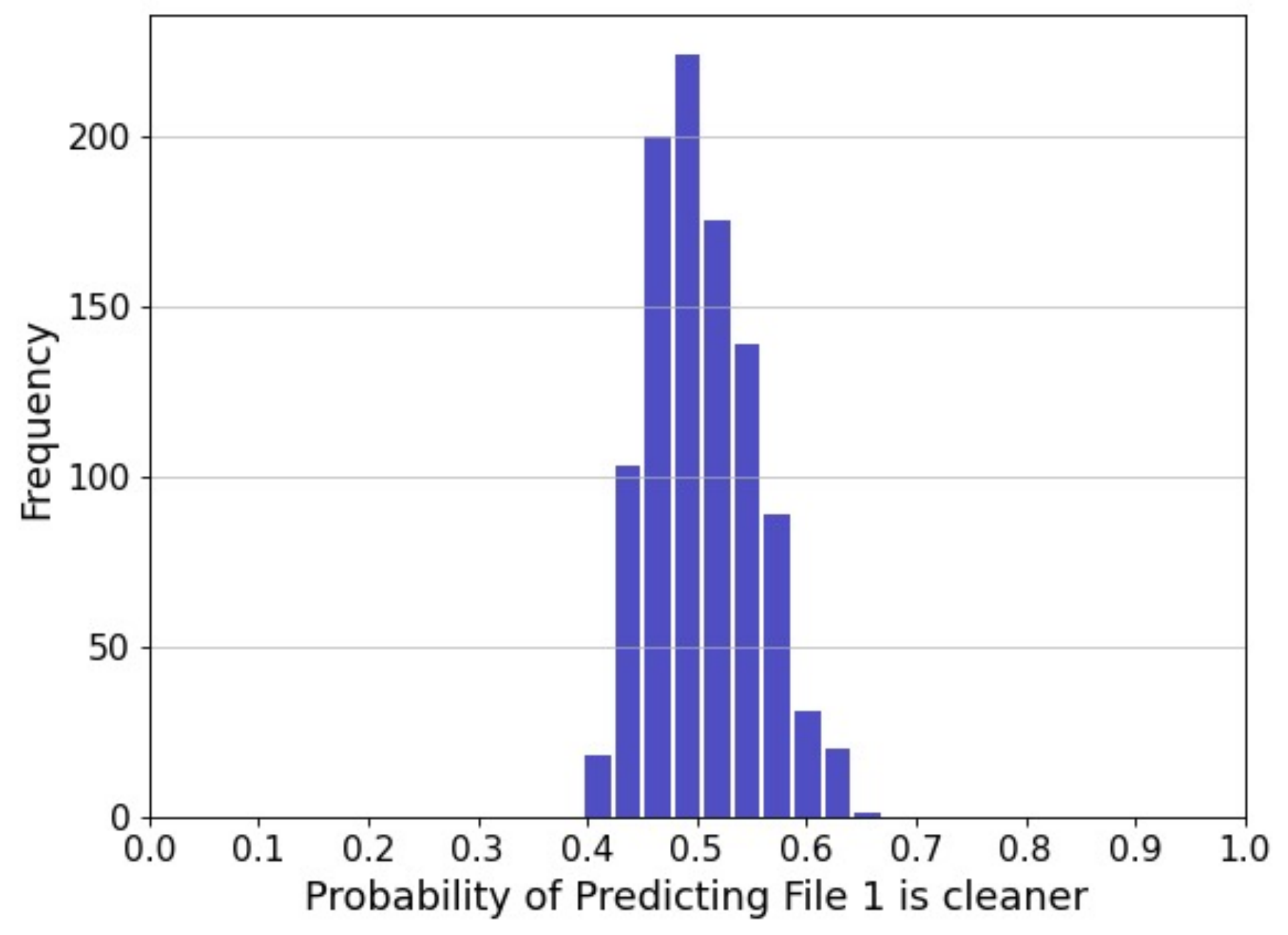} &
\includegraphics[width=\w]{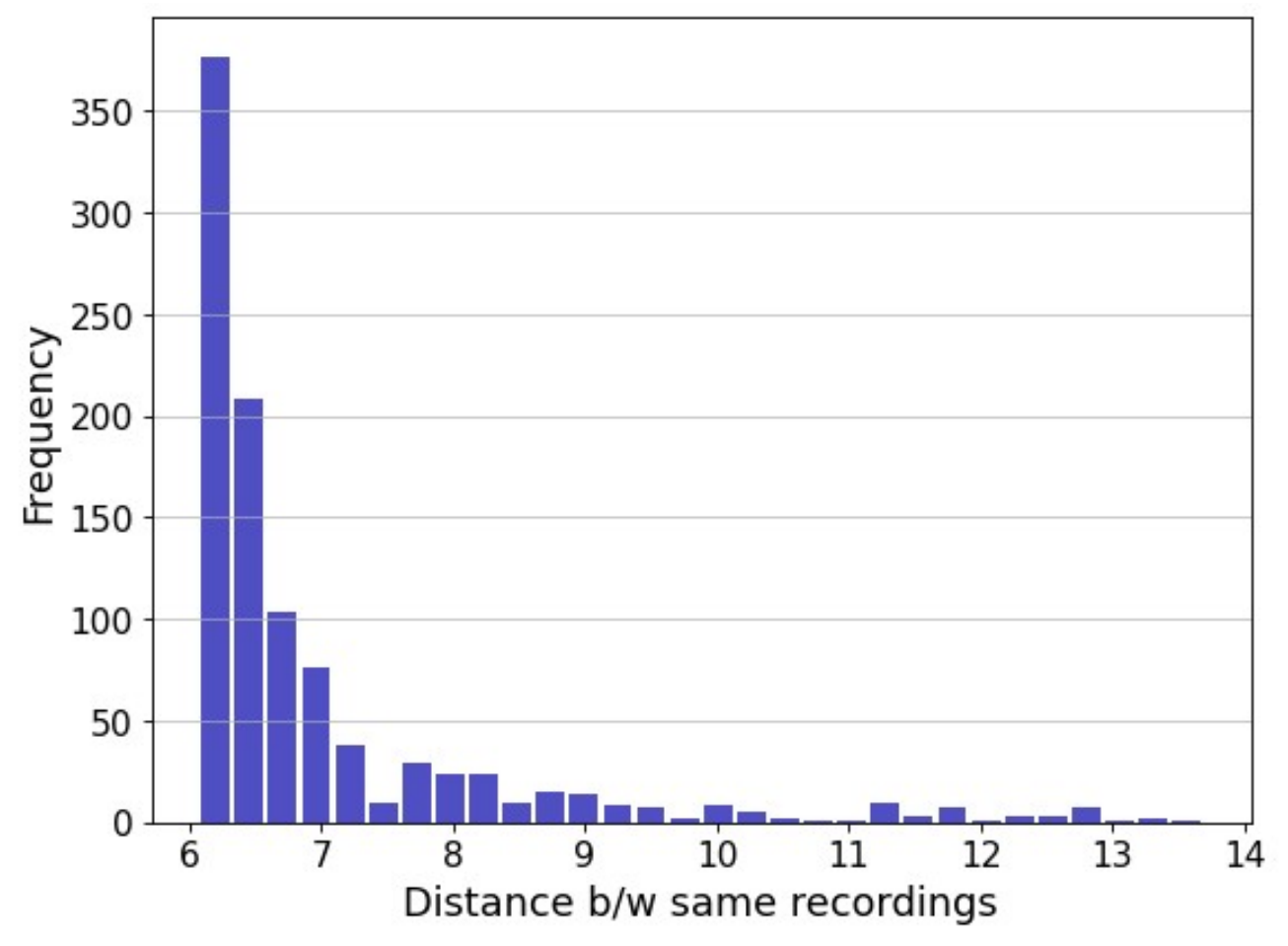} \\
\begin{minipage}{\w}
\centering
{\footnotesize (a)}

\end{minipage} &
\begin{minipage}{\w}
\centering
{\footnotesize (b)}

\end{minipage}
\end{tabular}
\caption{\textbf{Indiscernibility of Identicals}: (a)~Evaluating our metric's performance on the \textit{preference-task} and~(b) \textit{quantification-task} for the \textit{same} inputs}
\label{empirical_eval_f_x_x}
\end{figure}

Here, we show the output of our model when passed the same inputs (Refer to Sec 5.1 in the main paper). Ideally, the model should predict no quality difference when passed the same inputs. However, since our learning mechanism does not explicitly enforce the framework to have this property, so small errors are expected.
For a fair comparison, we calculate the recording-level predictions from our model.
Fig~\ref{empirical_eval_f_x_x}(a) shows the probability outputs on the \emph{preference-task}. The outputs are close to 0.5 that suggests that the model is unable to confidently identify which output is cleaner. 
Fig~\ref{empirical_eval_f_x_x}(b) shows the output from the \emph{quantification-task}. The distribution of the scores is centered around zero that suggests that the model correctly predicts that the two same inputs have similar quality levels, hence the near-zero quality difference.

\subsection{Framewise detection}

\begin{figure}[h!]
\centering
\includegraphics[width=0.50\columnwidth,]{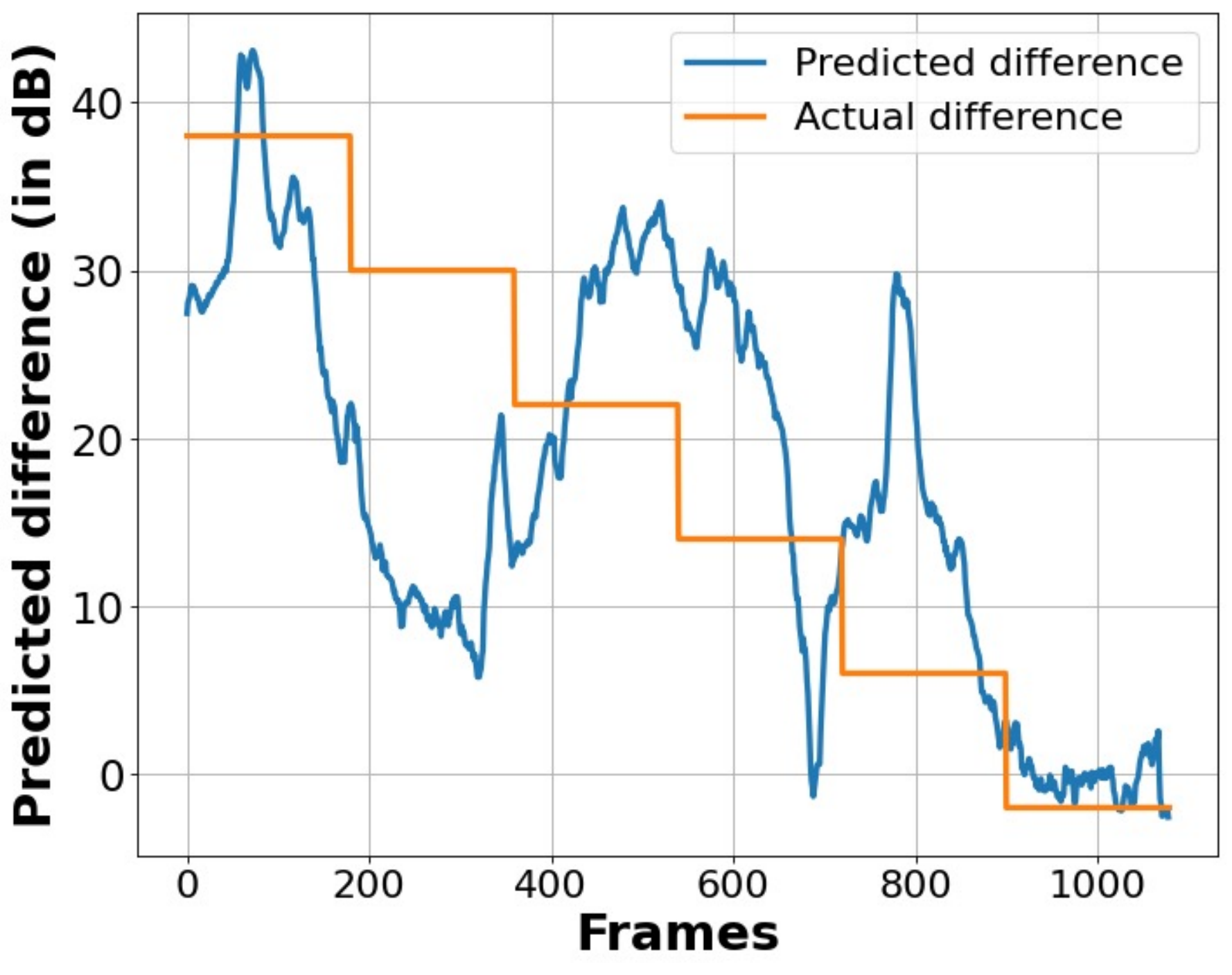}
\caption{{\bf Framewise Predictions:} Evaluating the framewise predictions of our model using a test recording that has decreasing levels of noise (from -10dB  to  +30dB) and the reference set consists of NMRs at +30dB.}
\label{framewise_detection}
\end{figure}

We also analyse the framewise performance of our model (Fig~\ref{framewise_detection}), using the same test bench that we created earlier, that consists of different recordings at various noise levels where each recording is 3 seconds. We concatenate different recordings together in \textit{decreasing} levels of noise (i.e from high noise to low noise equally spaced from -10dB to +30dB) which becomes the test input to our model. 
Equivalently, the reference input to our model consists of a random concatenation of NMRs at +30dB. 
We observe that the framewise output of task 1 (preference-task) is almost {\it 97\%} accurate in the preference task of predicting which frame is cleaner between the two input recordings, which suggests that it learns this quite well. We also compare the framewise outputs from task 2 (quantitative-task) that predicts their relative quality difference. We find that the framewise predictions are not monotonic but still follow the general trend. This is expected, since we always optimize on a recording level, and not on a frame level.
This findings suggest that our metric can detect and quantify which frames are degraded in quality between the inputs.

\section{Subjective Evaluation Datasets}

Refer to Sec 5.2 in the main paper. We evaluate our framework subjectively on two tasks:~(i) Correlation with MOS across 8 existing datasets, and~(ii) 2AFC accuracy, where we show the performance of our metric on triplet comparison questions from 4 different datasets. MOS checks for aggregated ordering, scale and consistency, whereas 2AFC checks for exact ordering of similarity at per sample basis. Most of the datasets lie within the range of -15dB to 60dB SNR and -15dB to 25dB SI-SDR, which also roughly matches our chosen training intervals.

The following are the datasets that we used for evaluation:
\begin{enumerate}[leftmargin=0.75cm]
  \item \textit{VoCo}:
  This dataset is based on comparing 6 different \emph{word synthesis and insertion} algorithms. The MOS tests were asked to rate which algorithm could synthesize a new word such that it blends seamlessly in the context of the existing narration. This consists of non-sample aligned pairs of recordings.
   
    \item \textit{Dereverberation}:
    This dataset is based on evaluating improvements across 5 deep-learning based \emph{speech enhancement} models including BLSTM, Wavenet, StarGAN-VC etc. MOS tests were done to evaluate which algorithm was rated the highest by subjects from Amazon Mechanical Turk (AMT). They obtained more than 100 ratings per condition.
   
  \item \textit{PEASS}: 
  The dataset contains separated sources and specifically defined anchor signals to assess audio \emph{source separation} performance across 4 metrics: \textit{global quality}, \textit{preservation of target source}, \textit{suppression of other sources}, and \textit{absence of additional artifacts}. Here, we only look at \textit{global quality}. It consists of scores from 20 subjects over a set of 80 sounds.
   
  \item \textit{Voice Conversion (VC)}: c
  This objective of this dataset is to compare the performance on speaker (voice) conversion. It consists of two tasks:~(i) parallel (\textit{HUB}) and~(ii) non-parallel data (\textit{SPO}). Here we only consider \textit{HUB}. The dataset consists of 4 source and 4 target speakers, where each speaker utters the same sentence set consisting of around 80 sentences.
   
  \item \textit{Noizeus}: 
  The dataset was developed to encourage comparison of non-deep learning based \emph{speech enhancement} algorithms, and consists of 30 IEEE sentences from 3 male and female speakers. The recordings are corrupted by 8 real-world noises, where the noises are taken from the AURORA database. The evaluation is done across 3 metrics: \textit{SIG}-speech signal alone; \textit{BAK}-background noise; and \textit{OVRL}-overall quality. Here, we only look at \textit{OVRL}.
   
  \item \textit{TCD-VoIP}: 
  This dataset was developed to help assess degradations that can occur in VoIP (Voice over IP calls). IT contains speech samples with a range of common VoIP degradations background noise, echo, chop (packet loss), clipping etc., and the corresponding set of subjective scores from 24 listeners.
   
  \item \textit{HiFi-GAN}:
  This dataset is based on evaluating the improvement across 10 deep-learning based \emph{speech enhancement} models including BLSTM, Wavenet, MetricGAN, SpecGAN etc. MOS tests were done to evaluate which algorithm was rated the highest by subjects from Amazon Mechanical Turk (AMT), with each method receiving around 14k ratings. 
  Further, it also consists of a 2AFC preference dataset consisting of around 1200 triplets, where each triplet received around 900 judgment ratings.
   
  \item \textit{FFTnet}:
  This dataset is based on evaluating the performance of 5 \emph{speech synthesis} algorithms across 2 male and female speakers. It introduces artifacts specific to synthesis and are not sample-aligned due to phase change.
  Further, it also consists of a 2AFC preference dataset consisting of around 2050 triplets, where each triplet received around 480 ratings.
   
  \item \textit{Bandwidth Expansion}:
      This dataset consists of subjective tests for 3 different \emph{bandwidth expansion} algorithms, aiming at increasing sample rate by filling in the missing high-frequency information. These audio samples consist of very subtle high-frequency differences. 
      Further, it also consists of a 2AFC preference dataset consisting of around 1020 triplets, where each triplet received around 400 judgments. 
    
    \item \textit{Simulated}:
    This dataset consists of 2AFC preference triplets, totalling to 1210. These are all based on adding \emph{common-realistic degradations} like background noises, speech distortions like clipping and other miscellaneous types of degradation's like compression and EQ.
   
\end{enumerate}

\section{Ablations}
In this section, we evaluate the influence of different components of our framework.

\subsection{Relative vs Absolute Quantification Task}
\label{subsec:absolute}
One key aspect of \frameworkname\, is that it tries to model relative quality differences rather absolute quality measures in terms of SNR and SI-SDR. We discussed motivations and intuitive reasons for it in the main paper. We empirically study the relative vs absolute modeling of the quality measures here. We compare our framework to models that are trained to predict SNR and SI-SDR directly. 

For the absolute quality prediction models, we consider two cases~(i) [\textit{Single Input Absolute Quantification}]: a single input model that takes a test recording and directly predicts the absolute quality measure , SNR and SI-SDR. It resembles the conventional formulation of a non-intrusive metric, and~(ii) [\textit{Two Input Absolute Quantification}]: a pairwise model that takes a test recording and a non-matching clean reference and predicts the absolute score. This is oriented towards our \frameworkname\, framework but instead of learning relative differences it tries to learn  absolute quality measures.

Results are shown in Table~\ref{table_mos_ablations_0}. We observe that our \emph{relative} quality prediction model~\frameworkname\, performs the best by a considerable margin. This empirically corroborate our hypothesis that learning to model relative differences is much better than absolute measures. Their correlations with MOS turns out to be much better than ``absolute methods". Moreover, we also observe that providing \emph{any} (even a non-matched) clean reference to the model improves the performance over the \textit{Single} case even for absolute quantification tasks. This demonstrates the inherent challenge of the conventional formulation of the non-intrusive metric that does not provide any reference. This highlights the usefulness of two features of our metric:~(i) predicting relative quality scores, and~(ii) providing non-matching references.



\begin{table}[h!]
\centering
      \setlength{\tabcolsep}{5pt}
      \resizebox{\columnwidth}{!}{
            \begin{tabular}{l l c c c c c c c c}
             \toprule
              \multirow{2}{*}{\bf Name} & \multirow{2}{*}{\bf Type} & \multicolumn{2}{c}{\bf VoCo} & \multicolumn{2}{c}{\bf Dereverb} & \multicolumn{2}{c}{\bf HiFi-GAN} & \multicolumn{2}{c}{\bf FFTnet} \\
              
             \cmidrule(lr){3-4} \cmidrule(lr){5-6} \cmidrule(lr){7-8} \cmidrule(lr){9-10}
             
             & & \bf PC &\bf SC & \bf PC & \bf SC & \bf PC & \bf SC & \bf PC & \bf SC \\
             \cmidrule(lr){1-10}
             
             \multirow{2}{*}{\bf Absolute}
              & {\bf Sing. Inp.} & 0.32 & 0.31 & 0.19 & 0.17 & 0.19 & 0.30 & 0.16 & 0.15 \\
             & {\bf Two Inp.} & 0.41$\pm$0.15 & 0.35$\pm$0.03 & 0.26$\pm$0.08 & 0.27$\pm$0.01 & 0.42$\pm$0.07 & 0.45$\pm$0.06 & 0.17$\pm$0.01 & 0.09$\pm$0.01  \\
             \cdashline{1-10}
             {\bf \frameworkname} &  & \bf 0.85$\pm$0.01 & \bf 0.68$\pm$0.03 & \bf 0.66$\pm$0.02 & \bf 0.67$\pm$0.02 & \bf 0.68$\pm$0.01 & \bf 0.78$\pm$0.01 & \bf 0.33$\pm$0.01 & \bf 0.44$\pm$0.01 \\
             \bottomrule
            \end{tabular}}
            \caption{\textbf{Ablations (1)}: Understanding the influence of predicting quality measures (single and pairwise) with our~\frameworkname\ using { $\textit{Global-Fixed}_{100}$} strategy. MOS Correlations: Spearman (SC), Pearson (PC). $\uparrow$ is better.}
            \label{table_mos_ablations_0}

\end{table}

\subsection{Multi-objective Learning of Quantification Task}

To evaluate the impact of the multi-objective optimization (i.e., optimizing over both SI-SDR and SNR) for the quantification task on correlation to subjective ratings, we compare our trained metric:~(i) only using the SNR head;~(ii) only using the SI-SDR head; and~(iii) after combining both SNR and SI-SDR heads. 
Results are shown in Table~\ref{table_mos_ablations_1}. For simplicity, we only look at the $\mbox{\textit{Unpaired-Global-Fixed}}_{100}$ strategy, which is evaluating a test recording using 100 clean NMRs randomly chosen from the DAPS dataset.
We observe that using either head alone performs worse than using both together, which suggests that using a multi-objective optimization helps learn a better general representation.

\begin{table}[h!]
\centering
      \setlength{\tabcolsep}{5pt}
      \resizebox{\columnwidth}{!}{
            \begin{tabular}{l l c c c c c c c c}
             \toprule
              \multirow{2}{*}{\bf Type} & \multirow{2}{*}{\bf Name} & \multicolumn{2}{c}{\bf VoCo} & \multicolumn{2}{c}{\bf Dereverb}& \multicolumn{2}{c}{\bf HiFi-GAN} & \multicolumn{2}{c}{\bf FFTnet} \\
              
             \cmidrule(lr){3-4} \cmidrule(lr){5-6} \cmidrule(lr){7-8} \cmidrule(lr){9-10}
             
             & &\bf PC &\bf SC & \bf PC & \bf SC & \bf PC & \bf SC & \bf PC & \bf SC \\
             \cmidrule(lr){1-10}
             \multirow{3}{*}{\bf \frameworkname\,}
             & {\bf SNR only} & 0.43 & 0.39 & 0.39 & 0.38 & 0.49 & 0.42 & 0.2 & 0.1  \\
             & {\bf SI-SDR only} & 0.6 & 0.48 & 0.48 & 0.49 & 0.54 & 0.65 & 0.25 & 0.28 \\
             & {\bf SNR and SI-SDR} & \bf 0.85 & \bf 0.68 & \bf 0.66 & \bf 0.67 & \bf 0.68 & \bf 0.78 & \bf 0.33 & \bf 0.44  \\
             \bottomrule
            \end{tabular}}
            \caption{\textbf{Ablations (2)}: Understanding the influence of using (multi-objective) SI-SDR and SNR for Task 2 using { $\textit{Global-Fixed}_{100}$} strategy. MOS Correlations: Spearman (SC), Pearson (PC). $\uparrow$ is better.}
            \label{table_mos_ablations_1}

\end{table}

\subsection{Number of NMRs}

Here we report the MOS correlation scores obtained when considering a set of 1, 10 and 100 NMRs for each test recording. 
This is shown for all 3 unpaired strategies described in Sec 5.2 (of the main paper) - \emph{Unpaired, Unpaired-Local-Fixed, and Unpaired-Global-Fixed}. Results are shown in Table~\ref{table_mos_3_ablation}. We observe that averaging the scores over a larger set of NMRs reduces the standard deviation in the scores which leads to more stable predictions. 
We also observe that PC values are more consistent and stable over many iterations, and have a lower standard deviation than SC values. Since SC measures monotonic relationships, it can easily overfit to a complex function, leading to a higher standard deviation per iteration. However, since PC maps linear relationships, it is more stable. 
Finally, we observe no significant difference between the scores from \textit{Unpaired-Local-fixed} and \textit{Unpaired-Global-fixed} which suggests that our metric works equally well for scenarios where we take \textit{any} random set of clean recordings as NMRs.

\begin{table}[h!]
\centering
\resizebox{\columnwidth}{!}{
 \begin{tabular}{l l c c c c c c c c}
 \toprule
  \multirow{2}{*}{\bf Type} & \multirow{2}{*}{\bf Category} & \multicolumn{2}{c}{\bf VoCo} & \multicolumn{2}{c}{\bf Dereverb}& \multicolumn{2}{c}{\bf HiFi-GAN}& 
 \multicolumn{2}{c}{\bf FFTnet}\\
 \cmidrule(lr){3-4} \cmidrule(lr){5-6} \cmidrule(lr){7-8} \cmidrule(lr){9-10}
 & &\bf PC &\bf SC & \bf PC & \bf SC & \bf PC & \bf SC & \bf PC & \bf SC\\
 \cmidrule(lr){1-10}
 \multirow{3}{*}{\bf Unpaired}
 
 & {\bf $\text{NMR}_1$} & 0.76$\pm$0.1 & 0.27$\pm$0.2 & 0.57$\pm$0.03 & 0.62$\pm$0.04 & 0.63$\pm$0.01 & 0.70$\pm$0.02 & 0.43$\pm$0.10 & 0.45$\pm$0.11 \\
 
 & {\bf $\text{NMR}_{10}$} & 0.87$\pm$0.01 & 0.43$\pm$0.07 & 0.64$\pm$0.01 & 0.73$\pm$0.03 & 0.63$\pm$0.01 & 0.70$\pm$0.01 & 0.45$\pm$0.03 & 0.48$\pm$0.06 \\
 
 & {\bf $\text{NMR}_{100}$} &  0.88$\pm$0.01 & 0.41$\pm$0.06 & 0.63$\pm$0.01	&  0.75$\pm$0.02 & 0.63$\pm$0.01	& 0.71$\pm$0.01 &	0.46$\pm$0.01 &	0.51$\pm$0.02 \\
 
 \cdashline{1-10}
 
 \multirow{3}{*}{\bf \hspace{0.005in}+Local-Fixed}
 & {\bf $\text{NMR}_1$} & 0.65$\pm$0.23 & 0.40$\pm$0.23
 & 0.53$\pm$0.10 & 0.57$\pm$0.15 & 0.56$\pm$0.08 & 0.64$\pm$0.08 & 0.38$\pm$0.10 & 0.31$\pm$0.13 \\
 
  & {\bf $\text{NMR}_{10}$} & 0.79$\pm$0.1 & 0.44$\pm$0.2 & 0.61$\pm$0.05 & 0.69$\pm$0.05 & 0.61$\pm$0.02 & 0.67$\pm$0.03 & \bf 0.48$\pm$0.03 & 0.50$\pm$0.04 \\

   & {\bf $\text{NMR}_{100}$} & \bf 0.89$\pm$0.01 & 0.44$\pm$0.06 & 0.63$\pm$0.01 & \bf 0.75$\pm$0.01 & 0.61$\pm$0.01 & 0.73$\pm$0.01 & 0.46$\pm$0.01 & \bf 0.51$\pm$0.02 \\

\cdashline{1-10}
\multirow{3}{*}{\bf \hspace{0.005in}+Global-Fixed}
 & {\bf $\text{NMR}_{1}$} & 0.79$\pm$0.20 & 0.54$\pm$0.20 & 0.44$\pm$0.16 & 0.41$\pm$0.19 & 0.56$\pm$0.08 & 0.63$\pm$0.10 & 0.29$\pm$0.10 & 0.36$\pm$0.12 \\

 & {\bf $\text{NMR}_{10}$} & 0.84$\pm$0.05 & 0.63$\pm$0.08 & 0.62$\pm$0.08 & 0.62$\pm$0.09 & 0.63$\pm$0.01 & 0.71$\pm$0.02 & 0.33$\pm$0.03 & 0.41$\pm$0.07 \\

 & {\bf $\text{NMR}_{100}$} & 0.85$\pm$0.01 & \bf 0.68$\pm$0.03 & \bf 0.66$\pm$0.02 & 0.67$\pm$0.02 & \bf 0.68$\pm$0.01 & \bf 0.78$\pm$0.01 & 0.33$\pm$0.01 & 0.44$\pm$0.02\\

 \bottomrule
\end{tabular}
}
\caption{\textbf{Ablations (3)}: Understanding the effect of number of recordings used as non-matching references (NMR). MOS Correlations: Pearson (PC),  Spearman (SC). Each cell shows the mean and standard deviation after 10 iterations. $\uparrow$ is better.}
\label{table_mos_3_ablation}

\end{table}

\section{Speech enhancement}

We use the VCTK dataset that consists of around 11,572 utterances for training and 824 files for validation. The dataset consists of 28 speakers equally split between male and female speakers, containing 10 unique background noise types across 4 different SNR conditions. 
Our denoising network is similar to Tan et al. and consists of a multi-layer convolutional encoder and decoder with U-Net skip connections, and a sequence modeling network applied on the encoders output. The input to the model are the real and imaginary components of the STFT of the signal, and the outputs are the complex ratio mask. The encoder (and decoder) consists of 5 gated convolutional layers with a filter size of 2 $\times$ 6, and use sigmoid activation for the gating mechanism. The sequence modeling network takes the encoders output and outputs a non-linear transformation of the same size. Since we design a \emph{causal} model, the network consists of 2 uni-directional LSTM layers with 256 hidden units in each layer.

For evaluation, we use the audio clips from the VCTK test set and evaluate scores on that dataset. We evaluate the quality of enhanced speech using objective measures. We use: i) PESQ (from 0.5 to 4.5); (ii) Short-Time Objective Intelligibility (\textit{STOI}) (from 0 to 100); (iii) Segmental Signal-to-Noise Ratio (\textit{SNRseg}): average of SNR values of short segments (15 to 20ms) ; (iv) \textit{CSIG}: MOS prediction of the signal distortion attending only to the speech signal (from 1 to 5); (v) \textit{CBAK}: MOS prediction of the intrusiveness of background noise (from 1 to 5); (vi) \textit{COVL}: MOS prediction of the overall effect (from 1 to 5). We compare the baseline approach with our model across the various paired-data constrained strategies.

As shown in Table 4 (main paper), SE models trained using~\frameworkname\,  obtain higher objective scores than the baseline models across all three strategies. We observe that the difference between the scores from our model and the baseline keeps increasing as we get more paired data which shows the usefulness of our metric, especially for sparse labeled-data situations (e.g., low-resource languages) since our approach leverages unlimited unpaired data.

Most notable is the improvement in \emph{STOI} that shows our metrics' utility as an optimization objective for pretraining. Given sparse-labeled data, the baseline model can fit the test set only as much. However, since our model can effectively leverage unlimited unpaired data during pretraining, given a model of sufficient capacity, it has the flexibility to learn a more complex mapping. This is precisely why we observe higher objective scores for quality and intelligibility than the baseline approaches which also highlights the usefulness of our framework.

\end{document}